\documentclass{aa}  

\usepackage{graphicx}
\usepackage{txfonts}
\newcommand{\hi } {{\rm H}\,{\small\rm I}}
\def\fl#1 {\textcolor{red}{#1}\;}
\usepackage[colorlinks=true, allcolors=blue]{hyperref}

\begin{document} 

   \title{Galaxy clusters in Milgromian dynamics: Missing matter, hydrostatic bias, and the external field effect}

    \titlerunning{Galaxy clusters in Milgromian dynamics}

   \author{R. Kelleher
          \inst{1, 2}
          \and
          F. Lelli\inst{2}
          }
    \institute{Dipartimento di Fisica e Astronomia “G. Galilei”, Università di Padova, vicolo dell’Osservatorio 3, I-35122 Padova, Italy
    \and
   INAF - Arcetri Astrophysical Observatory, Largo E. Fermi 5, 50125, Florence;
              \email{federico.lelli@inaf.it}}
            
   \date{Received 13 March 2024; accepted April 17, 2024}
    
 
  \abstract
   {We study the mass distribution of galaxy clusters in Milgromian dynamics, or modified Newtonian dynamics (MOND). We focus on five galaxy clusters from the X-COP sample, for which high-quality data are available on both the baryonic mass distribution (gas and stars) and internal dynamics (from the hydrostatic equilibrium of hot gas and the Sunyaev-Zeldovich effect). We confirm that galaxy clusters require additional `missing matter' in MOND, although the required amount is drastically reduced with respect to the non-baryonic dark matter in the context of Newtonian dynamics. We studied the spatial distribution of the missing matter by fitting the acceleration profiles of the clusters with a Bayesian method, finding that a physical density profile with an inner core and an outer $r^{-4}$ decline (giving a finite total mass) provide good fits within $\sim$1 Mpc. At larger radii, the fit results are less satisfactory but the combination of the MOND external field effect and hydrostatic bias (quantified as 10$\%$-40$\%$) can play a key role. The missing mass must be more centrally concentrated than the intracluster medium (ICM). For relaxed clusters (A1795, A2029, A2142), the ratio of missing-to-visible mass is around $1-5$ at $R\simeq200-300$ kpc and decreases to $0.4-1.1$ at $R\simeq2-3$ Mpc, showing that the total amount of missing mass is smaller than or comparable to the ICM mass. For clusters with known merger signatures (A644 and A2319), this global ratio increases up to $\sim$5 but may indicate out-of-equilibrium dynamics rather than actual missing mass. We discuss various possibilities regarding the nature of the extra mass, in particular `missing baryons' in the form of pressure-confined cold gas clouds with masses of $<10^5$ M$_\odot$ and sizes of $< 50$ pc.}

   \keywords{Cosmology: dark matter --  Galaxies: clusters: general -- Galaxies: clusters: intracluster medium --  Gravitation -- X-rays: galaxies: clusters}

   \maketitle
%

\section{Introduction}\label{sec:intro}

The nature of dark matter (DM) is one of the most puzzling mysteries in physics. A major alternative to non-baryonic DM is Milgromian dynamics, also known as modified Newtonian dynamics (MOND), which posits a departure from the standard non-relativistic laws of gravity and/or inertia below a characteristic acceleration scale of $a_0 \simeq 1.2\text{x}10^{-10}$ m s$^{-1}$ \citep{MOND2, MOND3, MOND1}. The MOND paradigm has been very successful on galaxy scales, reproducing the dynamical properties of different types of galaxies without the need for DM \citep[see][for reviews]{Famaey2012, Milgrom2014, Banik2022}. However, the main motivation of MOND is not merely to eliminate the need for DM, but rather to find a natural explanation for the dynamical regularities and scaling laws that galaxies obey \citep[e.g.][]{Lelli2017, McGaugh2020a}, some of which were indeed predicted by MOND before being observed \citep[see][for reviews]{McGaugh2020b, Lelli2022}. 

Moving beyond galaxy scales, the MOND paradigm is able to explain the dynamical properties of galaxy groups without the need for DM \citep{Milgrom2002, Milgrom2018, Milgrom2019}. In addition, basic MOND cosmological models predicted the existence of massive galaxies at $z\simeq10$ \citep{Sanders1998, Sanders2008} well in advance of their discovery by the James Webb Space Telescope; these still represent a challange for the $\Lambda$CDM cosmological model \citep[see][and references therein]{BoylanKolchin2023}. However, the development of a fully fledged MOND cosmology is still in progress and is concurrent with the construction of a proper relativistic extension of MOND. A recent example is the aether scalar tensor (AeST) theory proposed by \citet{relativisticmond}, which is able to reproduce the power spectrum of the cosmic microwave background (CMB), the linear mass power spectrum, and the equal speed of gravitational waves and electromagnetic waves.

A long-standing challenge for MOND appears at the intermediate scales of galaxy clusters \citep{The1988, Gerbal1992, Sanders1994, Sanders1999, Sanders2003, Aguirre2001, Pointecouteau2005, Takahashi2007, Natarajan2008, Angus2008, Ettori2019, Tian2020, Tian2021, Li2023}. As non-relativistic systems with internal accelerations of around $a_0$, galaxy clusters should behave as predicted by the basic MOND tenets. Instead, it has been clear for decades that MOND requires more mass in galaxy clusters than is observed, although the deficit is drastically reduced with respect to that implied by Newtonian dynamics. This missing mass cannot be contained inside individual galaxies but must be embedded within the intracluster medium (ICM), which largely dominates the known baryonic mass budget of galaxy clusters.

\begin{table*}
\centering
\caption{Properties of the cluster sample. $T_{\rm 3D}$ is the deprojected ICM temperature averaged across all probed radii, $R_{\rm out}$ is the outermost radius at which gravitational accelerations are measured using the SZ effect \citep[see][]{Eckert2019}, stellar mass ($M_\star$), gas mass ($M_{\rm gas}$), and baryonic mass ($M_{\rm bar}$) are measured within $R_{\rm out}$; their uncertainties are of the order of 25$\%$.}
\label{tab:properties}
\begin{tabular}{cccccccc}
 Cluster &  Redshift & Distance & $T_{\rm 3D}$ & $R_{\rm out}$ & $M_\star$ & $M_{\rm gas}$ & $M_{\rm bar}$  \\
& & (Mpc) & (keV) & (Mpc) & (10$^{14}$ M$_{\odot}$) & (10$^{14}$ M$_{\odot}$) & (10$^{14}$ M$_{\odot}$) \\
\hline
\multicolumn{8}{l}{\emph{Clusters without known merger signatures:}}\\
 A1795 & 0.0622 & 279.7 $\pm$ 19.7 & 3.27 $\pm$ 0.19 & 2.4 & 0.08 & 1.1 & 1.18 \\
 A2029 & 0.0766 & 350.7 $\pm$ 24.6 &5.75 $\pm$ 0.27& 2.7 & 0.22 & 2.2 & 2.42 \\
 A2142 & 0.0900 & 397.7 $\pm$ 27.9 &6.22 $\pm$ 0.27& 2.8 & 0.24 & 2.6  & 2.84 \\
\multicolumn{8}{l}{\emph{Clusters with known merger signatures:}}\\
A644 & 0.0704 & 315.3 $\pm$ 22.1 & 5.68 $\pm$0.22 & 1.7 & 0.09 & 1.0 & 1.09 \\
A2319 & 0.0557 & 243.7 $\pm$ 17.1 & 7.21 $\pm$ 0.14& 2.5 & 0.15 & 2.9 & 3.05 
\end{tabular}
\end{table*}

\citet{Sanders2003, Sanders2007} proposed that the MOND `missing mass' may consist of standard active neutrinos with masses of $\sim$2 eV, which were recently ruled out by the KATRIN experiment \citep{KATRIN1}. Other proposed solutions to the MOND galaxy cluster problem include: (1) undetected baryons, such as compact clouds of cold gas \citep{Milgrom2008}, (2) sterile neutrinos \citep{Angus2008, Angus2010}, (3) extended MOND theories in which the MOND acceleration scale varies with the depth of the gravitational potential \citep{Zhao2012, Hodson2017}, and (4) the effect of additional fields in relativistic extensions of MOND \citep{Durakovic2023}. Interestingly, cold gas clouds and/or sterile neutrinos could explain the Bullet Cluster in MOND similarly to active neutrinos \citep{Angus2007} because the missing mass would consist of non-collisional particles that become displaced from the hot gas during the clusters collision. Numerical calculations of colliding clusters in extended MOND theories have not yet been carried out.
In the present paper, we present a study of five galaxy clusters from the XMM-Newton cluster outskirts project \citep[X-COP,][]{Ghirardini2019}. The aims of this study are to constrain the spatial distribution of the MOND missing mass and to gain new insights into its possible nature. The X-COP sample represents the state of the art in terms of cluster data, providing high-quality X-ray imaging and spectroscopy for studying ICM properties, as well as optical imaging and spectroscopy that can be used to study the galaxy properties. In Sect.\,\ref{sec:data} we describe the cluster dataset, while in Sect.\,\ref{sec:methods} we present our fitting methodology. In Sect.\,\ref{sec:results}, the main fitting results are described.
In Sect.\,\ref{sec:disc} we discuss current observational constraints on the nature of the MOND missing mass, while in Sect.\,\ref{sec:sum} we provide a short summary of our findings. 

\section{Dataset}\label{sec:data}

\subsection{The cluster sample}\label{sec:sample}

We consider galaxy clusters from the X-COP project \citep{Ghirardini2019}, which measured the cluster gravitational field out to large radii combining X-ray data from XMM-Newton with Sunyaev-Zeldovich (SZ) data from Planck \citep{plancksz}. The X-COP clusters were selected to have (1) a signal-to-noise ratio (S/N) of higher than 12 in the SZ effect from Planck, (2) an expected halo masses of $M_{500} > 3 \text{×} 10^{14}M_{\odot}$, (3) a redshift in the range of $0.04<z<0.1$, (4) an apparent angular size of $\theta_{500} > 10$ arcmin, and (5) a hydrogen column density of $N_{\rm H} < 10^{21}$ cm$^{-2}$ along the line of sight. These criteria yielded 15 clusters; however, 3 clusters were excluded due to their complex morphologies, leaving a sample of 12 clusters \citep{Ghirardini2019}. Of these 12 clusters, five objects (A644, A1795, A2029, A2142, A2319) had additional measurements of the stellar mass distribution in the brightest cluster galaxy (BCG) and the surrounding satellite galaxies \citep{gravitationalxcop}. In this work, we focus on these five galaxy clusters with full baryonic information. Basic information on these galaxy clusters is given in Table \ref{tab:properties}. 

Importantly, the clusters A644 and A2319 show signatures of merger activity, so the hot gas may potentially be out of dynamical equilibrium. A2319 is a well-studied cluster undergoing a major merger \citep{Molendi1999, OHara2004, Govoni2004, Farnsworth2013, Yan2014, Storm2015}. A644 displays various indications of past and/or ongoing merger activity \citep{Bauer2000, Buote2005, Fusco2009}. We keep these two clusters in our sample for the sake of comparison, but we stress that they may not be appropriate to test a dynamical theory such as MOND.

 \subsection{Observed gravitational field} \label{sec:obs}

Assuming that the hot gas is a spherically symmetric system in hydrostatic equilibrium, the pressure gradient exactly balances the gravitational attraction, so that
\begin{equation}\label{eq:hydro}
     g_{\rm obs}(r) = -\frac{1}{\rho_{\rm gas}(r)} \frac{dP_{\rm gas}(r)}{dr},
\end{equation}
where $g_{\rm obs}$ is the observed gravitational acceleration, $P_{\rm gas}$ is the gas pressure, and $\rho_{\rm gas}$ is the gas volume density. X-ray imaging provides the emissivity $\epsilon_{\rm \nu}$ that is proportional to $n_{\rm e}^2 \propto \rho_{\rm gas}^2$, where $n_{\rm e}$ is the electron density of the ionised gas. To infer $\rho_{\rm gas}$, one typically assumes no gas clumping and spherical geometry in order to deproject line-of-sight integrated quantities into 3D radial quantities. In addition, X-ray spectroscopy provides the gas temperature profile $T_{\rm gas}$, again after appropriate emission-weighted deprojections \citep[see, e.g.][for details]{gravitationalxcop}. Assuming the ideal gas law, we have
\begin{equation}\label{eq:idealgas}
    P_{\rm gas}(r) = \frac{k_{\rm B}}{\kappa m_{\rm p}} \rho_{\rm gas}(r) T_{\rm gas}(r),
\end{equation}
where $k_{\rm B}$ is Boltzmann's constant, $m_{\rm p}$ is the proton mass, and $\kappa$ is the mean atomic weight of the gas that depends on its chemical composition ($\kappa=0.6$ for fully ionised gas with solar abundances). Eq.\,\ref{eq:idealgas} allows measuring the gas pressure profile and calculating its radial derivative to infer $g_{\rm obs}(r)$ from Eq.\,\ref{eq:hydro}. For numerical reasons, it is convenient to describe $\rho_{\rm gas}(r)$ and $T_{\rm gas}(r)$ via the sum of $N$ analytic functions with a set of free parameters $\vec{\alpha}$, so that the derivative $dP_{\rm gas}/dr$ can be calculated analytically once the values of $\vec{\alpha}$ are specified \citep[see][]{gravitationalxcop}. 

At large cluster radii, the gas pressure can also be measured from the SZ effect through a unit-less Comptonization parameter $y$, which is given by
\begin{equation}\label{eq2.1}
    y = \frac{\sigma_T}{m_e c^2} \int_l P_{\rm gas}(l) dl,
\end{equation}
where $\sigma_T$ is the Thomson cross section, $m_e$ the mass of the electron, and $c$ is the speed of light. Basically, the SZ effect provides the gas pressure $P_{\rm gas}$ integrated along the line of sight $l$, so the 3D pressure profile can be determined by deprojecting the $y$-profiles assuming spherical symmetry. In this work, we use the radial profiles of $\rho_{\rm gas}$, $T_{\rm gas}$, $P_{\rm gas}$, and $g_{\rm obs}$ computed by \citet{gravitationalxcop} using the so-called `non-parametric log-normal mixture reconstruction'. This is the most empirical, model-independent method to compute $g_{\rm obs}(r)$.

In Newtonian dynamics, one equates $g_{\rm obs}$ to the Newtonian gravitational field given by
\begin{equation}\label{eq:Newton}
g_{\rm N}(r) = \dfrac{G_{\rm N} M_{\rm tot}(<r)}{r^2},
\end{equation}
where $G_{\rm N}$ is Newton's constant and $M_{\rm tot}$ is the `total' dynamical mass, which is generally much larger than the visible baryonic mass $M_{\rm bar}$, implying large amounts of DM. The total mass profile can then be measured combining Eq.\,\ref{eq:hydro}, \ref{eq:idealgas}, and \ref{eq:Newton}. 

In Milgromian dynamics, it is most natural to work in terms of accelerations, so that
\begin{equation}\label{eq:mond}
    g_{\rm MOND} \mu \left(\frac{g_{\rm MOND}}{a_0}\right) = g_{\rm N, bar},
\end{equation}
where $g_{\rm N, bar}$ is the Newtonian baryonic gravitational field (discussed in the next section), $g_{\rm MOND}$ is the expected MOND acceleration to be equated to $g_{\rm obs}$, and $\mu(x)$ is the MOND interpolating function with $x=g_{\rm MOND}/a_0$. The functional form of $\mu(x)$ is not specified by the general MOND paradigm, but its asymptotic limits must be
\begin{equation}
    \mu(x \gg 1) \approx 1 \quad \rm{and}\quad
    \mu(x \ll 1) \approx x. 
\end{equation}
In particular, the low acceleration limit can be derived assuming scale invariance after the empirical normalisation of $a_0$ \citep{Milgrom2009}. Eq. \ref{eq:mond} is strictly valid in MOND modified gravity theories for isolated, spherically symmetric systems \citep{Bekenstein1984, Milgrom2010, Milgrom2023}. Eq. \ref{eq:mond} is also valid in MOND modified inertia theories in the case of isolated systems with purely circular orbits \citep{Milgrom1994}, which is clearly not the case for the random gas motions in the ICM. In this work, therefore, we are primarily testing MOND modified gravity theories, albeit we expect Eq.\,\ref{eq:mond} to provide the correct order of magnitude also in modified inertia theories (for isolated systems).

\subsection{Newtonian baryonic gravitational field}\label{sec:bar}

The Newtonian gravitational field sourced by baryons ($g_{\rm N, bar}$) is given by the contributions from the ICM ($g_{\rm N, ICM}$) and from cluster galaxies ($g_{\rm N, gal}$). The contributions from globular clusters and intracluster light (made of free-floating stars inside the galaxy cluster) are negligible. At large cluster radii, the majority ($\sim$90\%) of the baryonic contribution is from the ICM, while a minor fraction ($\sim$10\%) is from gas and stars inside galaxies. At small radii, however, the BCG contribution can be important. 

In this work, we use the values of $g_{\rm N, bar}=g_{\rm N, ICM}+g_{\rm N, gal}$ provided by \citet{gravitationalxcop}. The values of $g_{\rm N, ICM}$ are computed using $\rho_{\rm gas}$ from X-ray imaging. In particular, the observed emissivity profile $\epsilon_\nu(r)\propto\rho_{\rm gas}^2(r)$ is described as the sum of several King functions, for which the correspondence between projected 2D profiles and intrinsic 3D ones can be written analytically, so the deprojection is trivial \citep[see][]{Eckert2020}. The radial profile of $g_{\rm N, gal}$, instead, consists of two different components: the central BCG ($g_{\rm N, BCG}$) and satellite galaxies ($g_{\rm N, sat}$). Their derivation is briefly summarised in the following.

The mass distribution of the BCG is measured using $r$-band imaging and assuming a radially constant mass-to-light ratio $\Upsilon_*$. The value of $\Upsilon_*$ was dynamically estimated by \citet{Loubser2020} by modelling the observed stellar kinematics. In particular, the observed stellar kinematics was fitted considering - in addition to the stellar contribution - a supermassive black hole and a DM halo, whose mass was determined using weak lensing data. Ideally, in a MOND context, one should redo the stellar kinematic fits without a DM halo using Milgromian dynamics, but the final values of $\Upsilon_*$ are expected to be similar within the uncertainties because the DM contribution is small in the central parts of BCGs (in  MONDian terminology, the galaxy is in the Newtonian regime at $a\gg a_0$). Moreover, the uncertainties on $g_{\rm N, BCG}$ due to the DM halo parameters are included in the error budget \citep[see][for details]{gravitationalxcop}. For simplicity, therefore, we use the values computed by \citet{Loubser2020}. 

The mass distribution of satellite galaxies was computed using u-, g-, r- and i- band imaging; the values of $\Upsilon_*$ were estimated from spectral energy distribution fitting with stellar population models \citep[see][for details]{vanderBurg2015}. The gas content inside satellite galaxies is neglected, but it is expected to be very small (smaller than the uncertainties on $\Upsilon_*$) in passive cluster galaxies dominated by old stellar populations.

The Newtonian baryonic gravitational field is given by 
\begin{equation}\label{eq:gbar}
g_{\rm N, bar} = g_{\rm N, ICM} + g_{\rm N, BCG} + g_{\rm N, sat} 
\end{equation}
Any additional `missing mass' component could be trivially added to Eq.\,\ref{eq:gbar}. Since we are assuming spherical symmetry, the Newtonian enclosed mass profile and the corresponding circular-velocity curve of a component `i' are given by
\begin{equation}\label{eq:shelltheorem}
    M_{\rm i}(<r) = \frac{g_{\rm N, i}(r) \cdot r^2}{G_{\rm N}} \quad \rm{and} \quad V_{\rm N,i}(r) = \sqrt{g_{\rm N, i}(r) \cdot r}.
\end{equation}

\begin{table*}
\centering
\caption{Estimated parameters from MCMC fits (see Sect.\,\ref{sec:fit} for details). The ratio $M_{\rm mm}/M_{\rm bar}$ is measured at $R_{\rm out}$ (see Table\,\ref{tab:properties}). The parameters of A644 and A2319 should be interpreted with caution due to possible out-of-equilibrium effects.}
\begin{tabular}{cccccccc}
Cluster & Model & $\log(\Upsilon_{\rm bar})$& $\log(M_{\rm mm, tot}/M_\odot)$ & $\log(r_{\rm s}/{\rm kpc})$ & $\rho_0  ({\rm 10^{-25} g\,cm^{-3}})$&  $g_{\rm Ne}/{a_0}$ & $M_{\rm mm}/M_{\rm bar}$\\
\hline

 \multicolumn{8}{l}{\emph{Clusters without known merger signatures}}\\
A1795 & with EFE & -0.01$\pm$ 0.08 & 14.17$\pm$ 0.05 &  2.12 $\pm$ 0.02 & 10.53 $\pm$ 1.75  & 0.0025 $^{+0.0020}_{-0.0021}$& 1.10 $\pm$ 0.10 \\
 & no EFE & 0.07 $\pm$ 0.07 & 14.09 $\pm$ 0.04 & 2.09 $\pm$ 0.02 & 10.59 $\pm$ 1.74 & ... & 0.76 $\pm$ 0.09 \\
A2029 & with EFE &  0.07 $^{+0.05}_{-0.06}$& 14.31$\pm$ 0.04 & 2.17 $\pm$ 0.02 & 10.25 $\pm$ 1.70 & 0.0009 $^{+0.0006}_{-0.0007}$ & 0.61 $\pm$ 0.09 \\
 & no EFE & 0.19$^{+0.05}_{-0.06}$ & 14.21 $\pm$ 0.05 & 2.15 $\pm$ 0.02 & 9.24 $\pm$ 1.81 & ... & 0.37$\pm$0.08 \\
A2142 & with EFE & 0.23 $^{+0.05}_{-0.06}$ & 14.36$^{+0.07}_{-0.08}$& 2.39 $\pm$0.04 &  2.49 $\pm$ 0.80 & 0.0015 $\pm $0.0013  & 0.38
 $\pm$ 0.096 \\
 & no EFE &  0.28 $^{+0.05}_{-0.06}$ & 14.24 $^{+0.09}_{-0.11}$ &  2.37 $\pm$ 0.05 & 2.25 $\pm$ 0.88 &...&  0.26 $\pm$ 0.084 \\ 
 \multicolumn{8}{l}{\emph{Clusters with known merger signatures}}\\
A644 & with EFE & -0.14$^{+0.08}_{-0.09}$ & 14.78$\pm$ 0.05 & 2.40 $\pm$ 0.02 & 6.06 $\pm$ 1.12& 0.11 $\pm0.05$  & 5.43 $\pm$ 0.41 \\
 & no EFE & -0.11 $\pm$ 0.09 & 14.59 $\pm$ 0.03 & 2.34 $\pm$ 0.02 & 5.88 $\pm$ 0.86 & ... &3.38
 $\pm$ 0.18 \\
A2319 & with EFE & -0.21$^{+0.07}_{-0.08}$ &  15.05$\pm$ 0.03 & 2.59 $\pm$ 0.01 &  3.06 $\pm$ 0.37  & 0.29 $\pm$0.10 & 3.99 $\pm$ 0.11 \\
 & no EFE &  0.00$\pm$ 0.07 &  14.75 $\pm$ 0.03 &  2.54 $\pm$ 0.02 & 2.29 $\pm$ 0.28 & ...& 1.29 $\pm$ 0.04 \\ 
\end{tabular}
\label{tab:fit}
\end{table*}

\section{Methodology}\label{sec:methods}

\subsection{MOND theoretical formalism}\label{sec:fit}

The best approach to constrain the distribution of the missing matter in MOND is to assume a parametric density profile and directly fit it to the data, as is traditionally done for DM halos in Newtonian dynamics \citep[e.g.][]{Li2020}. This method is more robust to uncertainties than the subtraction approach used by \citet{gravitationalxcop}, which we revisit in Appendix\,\ref{sec:sub}.

For practical reasons, it is convenient to rewrite the MOND law for isolated systems (Eq.\,\ref{eq:mond}) as follows:
\begin{equation}\label{eq:nu}
    g_{\rm MOND, iso} = \nu \left(\frac{g_{\rm N}}{a_0}\right) g_{\rm N},
\end{equation}
where $\nu(y)\cdot\mu(x) = 1$ with $x=g_{\rm MOND}/a_0$ and $y=g_{\rm N}/a_0$. Eq.\,\ref{eq:nu} is mathematically equivalent to the radial acceleration relation of galaxies \citep{Lelli2017}, but it is important to stress that the former is a MOND prediction that applies in specific situations, while the latter is an empirical description of the observed dynamics of galaxies. For example, Eq.\,\ref{eq:nu} does not apply in cases where the so-called external field effect (EFE) is relevant.

The EFE is a characteristic prediction of MOND due to the violation of the strong equivalence principle (SEP). In general relativity, the SEP dictates that the internal dynamics of a gravitational system are independent of any external gravitational field in which the system is embedded (apart from possible tidal forces). MOND breaks the SEP due to its non-linearity \citep{Bekenstein1984}, while preserving the weak equivalence principle (the universality of free fall). Thus, in MOND, the internal dynamics of a gravitational system can depend on its location in space and time due to the EFE.

In general, accounting for the EFE requires complex numerical 3D computations that solve the MOND modified Poisson equation \citep[e.g.][]{Chae2022}. To a first order approximation, the EFE can be analytically calculated considering a 1D solution in which the internal and external accelerations are summed in modulus as if their vectors always have the same direction. This gives the formula \citep{Famaey2012}:
\begin{equation}\label{eq:nuEFE}
    g_{\rm MOND, EFE} = g_{\rm N} \nu\left( \dfrac{g_{\rm N} + g_{\rm N,e}}{a_0} \right) + g_{\rm N, e} \left[ \nu\left( \dfrac{g_{\rm N} + g_{\rm N,e}}{a_0}\right) - \nu\left( \dfrac{g_{\rm N,e}}{a_0}\right)\right]
\end{equation}
where $g_{\rm N, e}$ is the Newtonian external field from the large-scale distribution of baryonic mass in the Universe. 

In MOND modified gravity theories, Eq.\,\ref{eq:nuEFE} is an approximated formula that maximises the EFE because $g_{\rm N}$ and $g_{\rm N, e}$ are summed in modulus rather than in a vectorial way. Eq.\,\ref{eq:nu} and Eq.\,\ref{eq:nuEFE}, therefore, provide two extreme MOND scenarios: no EFE and maximal EFE. Any other EFE implementation is expected to give intermediate results. A possible exception is represented by MOND modified inertia theories in which the EFE is effectively given by several times the instantaneous value of $g_{\rm N, e}$ \citep{Milgrom2022}, so Eq.\,\ref{eq:nuEFE} may possibly underestimate the EFE in those specific theories.

In Eq.\,\ref{eq:nu} and Eq.\,\ref{eq:nuEFE}, we set
\begin{equation}
g_{\rm N}(r) = g_{\rm N, mm}(r; M_{\rm mm, tot}, r_{\rm s}) + \Upsilon_{\rm bar} g_{\rm N, bar}(r),
\end{equation}
where $g_{\rm N, mm}$ is the Newtonian gravitational field of the missing mass component, which depends on two free parameters ($M_{\rm mm}, r_{\rm s}$) as we describe in Sect.\,\ref{sec:prof}. The quantity $\Upsilon_{\rm bar}$, instead, is a nuisance parameter that scales up or down the baryonic component, similarly to a baryonic mass-to-light ratio. The purpose of $\Upsilon_{\rm bar}$ is to account for systematic uncertainties in $g_{\rm bar}$ (or equivalently in the total baryonic mass), which may be due to uncertainties in absolute X-ray flux calibration, deprojection methodology, gas clumpiness, assumed mean atomic weight, and so on. We impose a lognormal prior on $\log_{10}(\Upsilon_{\rm bar})$ in a Bayesian context, centered at 0 with a standard deviation of 0.1 dex, corresponding to average systematic uncertainties in the measured baryonic mass of $\sim$25$\%$.

\subsection{Bayesian fitting formalism}\label{sec:fit}

The fitting parameters are determined using a Markov-Chain-Monte-Carlo (MCMC) method in Bayesian statistics. We define the likelihood $\mathcal{L} = \exp({-0.5 \chi^2})$ with
\begin{equation}\label{eq:chi}
 \chi^2 = \sum_{k}^{N} \frac{[g_{\rm obs} - g_{\rm MOND}(\vec{p})]^{2}}{\delta^{2}_{g_{\rm obs}}},
\end{equation}
where $g_{\rm obs}$ is the observed acceleration (Eq.\,\ref{eq:hydro}) at radius $R_k$, $\delta_{g_{\rm obs}}$ is the associated error, and $g_{\rm MOND}$ is given by Eq.\,\ref{eq:nu} or Eq.\,\ref{eq:nuEFE}. The model acceleration depends on the fitting parameters $\vec{p}=\{ M_{\rm mm}, r_{\rm s}, \Upsilon_{\rm bar} \}$ plus $g_{\rm N, e}$ in the EFE case. The posterior probability distributions of $\vec{p}$ are mapped using \texttt{emcee} \citep{Foreman2013}. The MCMC chains are initialised with 200 walkers. We run 1000 burn-in iterations, then the sampler is run for another 2000 iterations. The \texttt{emcee} parameter $a$, which controls the size of the stretch move, is set equal to 2. This generally gives acceptance fractions larger than 50\%. In all cases, the MCMC posterior probability distributions are well-behaved and show a single peak (see Appendix\,\ref{app:corner}), indicating that the fitting parameters are well determined. The maximum-likelihood values and their 68\% confidence intervals (1$\sigma$ errors) are given in Table\,\ref{tab:fit}.

At large cluster radii ($r\gtrsim1$ Mpc), an important systematic uncertainty is the so-called `hydrostatic bias', which we discuss in detail in Sect.\,\ref{sec:hydrobias}. In short, the outer gas may not be in hydrostatic equilibrium, so Eq.\,\ref{eq:hydro} may not provide a robust measurement of $g_{\rm obs}$. We aim to estimate the possible amount of hydrostatic bias in MOND, but its effect is degenerate with the strength of the EFE (see also Appendix\,\ref{sec:sub}). To have upper and lower limits on hydrostatic bias in MOND, we proceed as follows. In the no-EFE case (Eq.\,\ref{eq:nu}) we fit only the data at $r<1$ Mpc, so that we obtain the maximum amount of hydrostatic bias needed in MOND. In the EFE case (Eq.\,\ref{eq:nuEFE}) we fit all data points, so that we estimate the minimum amount of hydrostatic bias needed in MOND. The fiducial radius of 1 Mpc is empirically inferred using the basic subtraction approach described in Appendix\,\ref{sec:sub}.

\subsection{Density profiles for the missing mass}\label{sec:prof}

We explored various density profiles for the missing mass and found good results with cored profiles of the following type:
\begin{equation}\label{eq:rho_mm}
    \rho_{\rm mm}(r) = \frac{\rho_0}{(1+\frac{r}{r_s})^n}
\end{equation}
where $\rho_0$ is the central core density, $r_s$ is a scale radius, and $n$ is the outer slope of the density profile. We choose density profiles with a denominator of the type $(1 + \mathcal{R})^n$ rather than $(1 + \mathcal{R}^n)$ for the sake of simplicity (where $\mathcal{R}=r/r_{\rm s}$). In the former case, indeed, the enclosed mass can be expressed by simple analytic formulas for integer values of $n$ (see, e.g. Eq.\,\ref{eq:M_n4}), which is not true in the latter case for most values of $n$. For a fixed value of $n$, we do not expect major differences between these two types of cored profiles. Notably, a cored profile is expected in the case that the missing mass is made of massive neutrinos \citep{Sanders2007}.

We also tested truncated spheres for which $\rho_{\rm mm}(r)=0$ at $r>r_{\rm t}$, where $r_{\rm t}$ is a truncation radius. We explored different behaviours of $\rho_{\rm mm}(r)$ at $r<r_{\rm t}$, such as $\rho_{\rm mm}(r)=\rm{const}$, $\rho_{\rm mm}(r)\propto r$, and $\rho_{\rm mm}(r) \propto 1/r$, but in all cases we found that a truncated sphere did not improve the fits with respect to Eq.\,\ref{eq:rho_mm}. 

In Newtonian dynamics, to obtain a flat velocity curve, the mass density profile must have an outer slope $n\simeq2$ (e.g. an isothermal sphere). A value of $n=2$ gives an enclosed mass profile that linearly increases with $r$ and unphysically diverges to infinity at large radii. In $\Lambda$CDM cosmology, DM halos are predicted to have $n\simeq2$ only at intermediate radii, while they tend to $n\simeq3$ at large radii (e.g. the NFW profile), so the enclosed DM mass logarithmically diverges to infinity. For $n>3$, instead, the enclosed mass converges to a physical, finite value.
We explored various values of $n$ (from 2 to 6) and found that $n=4$ provides the best fits (but we note that the Bayesian Information Criterion provides very small differences for fits with $n\ge4$). An outer slope $n=4$ is particularly interesting because it corresponds to that expected for a MOND isothermal sphere \citep{Milgrom1984}. For $n=4$, the total enclosed mass is
\begin{equation}\label{eq:M_n4}
    M_{\rm mm}(<r) = \int_0 ^{r} 4 \pi r'^2 \frac{\rho_0}{(1+\frac{r'}{r_s})^4} dr' =   \frac{4 \pi \rho_0 r_s^3}{3} \frac{\big(\frac{r}{r_s}\big)^3}{\big(1+\frac{r}{r_s}\big)^3}.
\end{equation}
In the limit of $r \rightarrow \infty$, the mass converges to
$M_{\rm mm, tot} = 4/3 \pi \rho_0 r_s^3$, so it is convenient to rewrite Eq.\,\ref{eq:M_n4} as
\begin{equation}
    M_{\rm mm}(\mathcal{R}) = M_{\rm mm, tot}\frac{\mathcal{R}^3}{(1+\mathcal{R})^3}.
\end{equation}
with the dimensionless variable $\mathcal{R}=r/r_{\rm s}$. The Newtonian gravitational field of the missing mass component ($g_{\rm mm}$) is then given by Eq.\,\ref{eq:Newton} and depends on two fitting parameters $(M_{\rm mm, tot}, r_{\rm s})$.

\begin{figure*}
\centering		 
	\includegraphics[width=4.5cm, height=4.5cm]{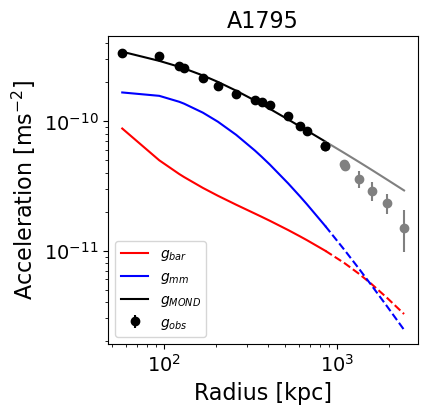}
	\includegraphics[width=4.5cm, height=4.5cm]{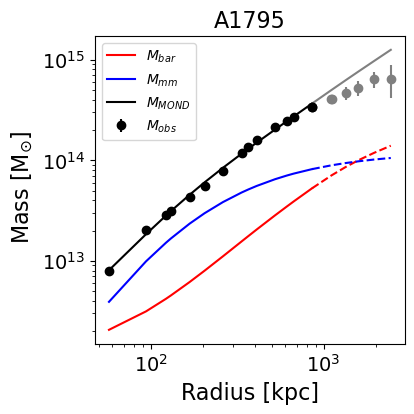}
	\includegraphics[width=4.5cm, height=4.5cm]{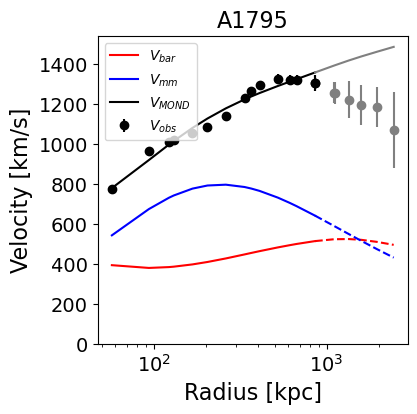}
	\includegraphics[width=4.5cm, height=4.5cm]{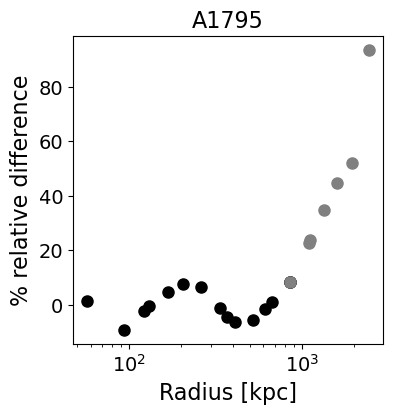}
          
	\includegraphics[width=4.5cm, height=4.5cm]{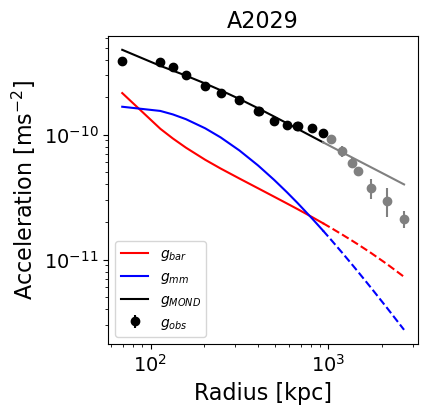}
	\includegraphics[width=4.5cm, height=4.5cm]{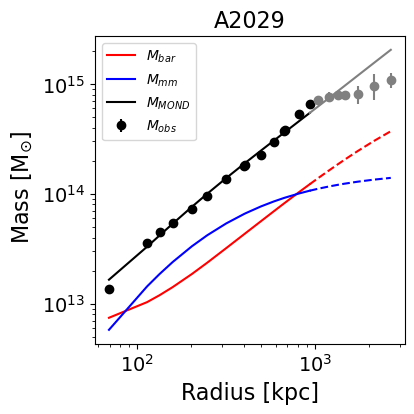}
	\includegraphics[width=4.5cm, height=4.5cm]{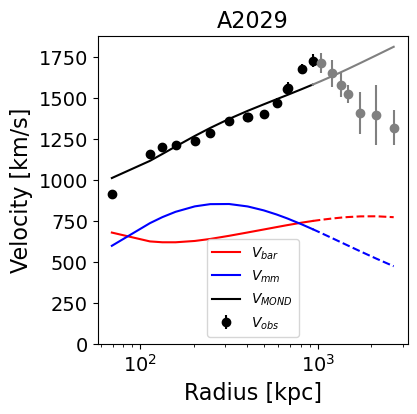}
	\includegraphics[width=4.5cm, height=4.5cm]{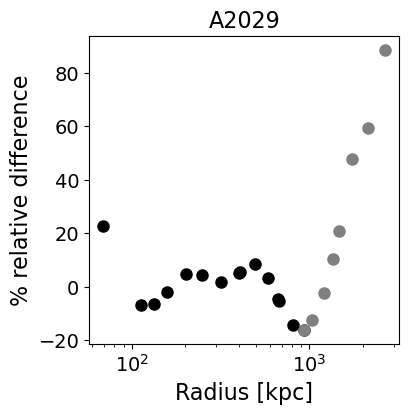}
          
	\includegraphics[width=4.5cm, height=4.5cm]{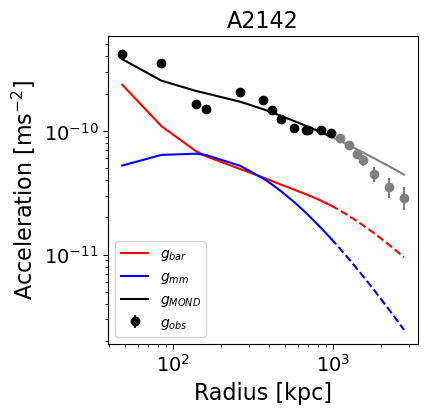}
	\includegraphics[width=4.5cm, height=4.5cm]{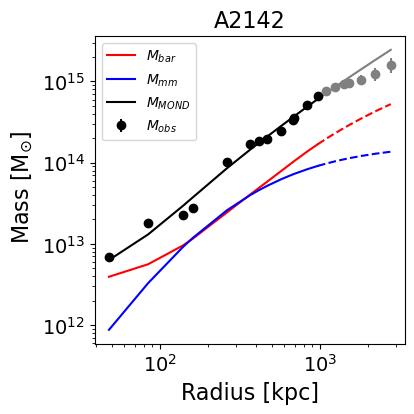}
	\includegraphics[width=4.5cm, height=4.5cm]{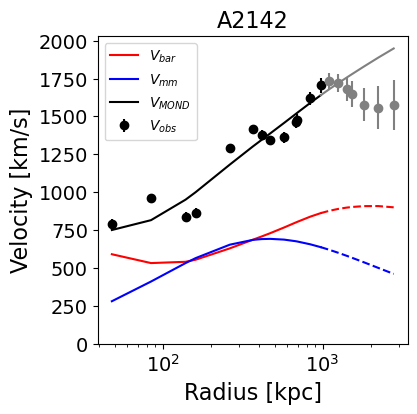}
	\includegraphics[width=4.5cm, height=4.5cm]{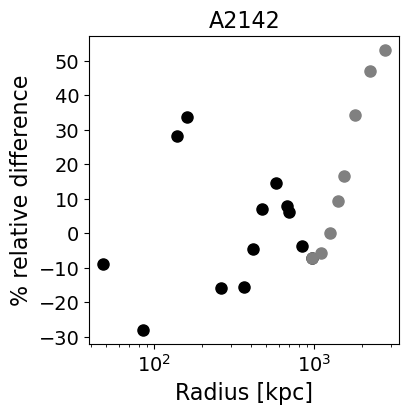}

    \includegraphics[width=4.5cm, height=4.5cm]{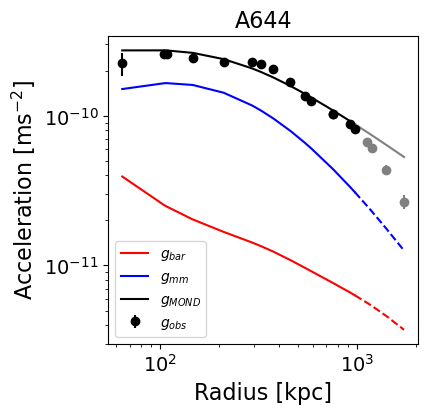}
	\includegraphics[width=4.5cm, height=4.5cm]{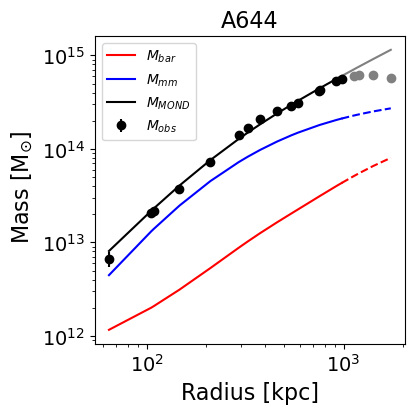}
	\includegraphics[width=4.5cm, height=4.5cm]{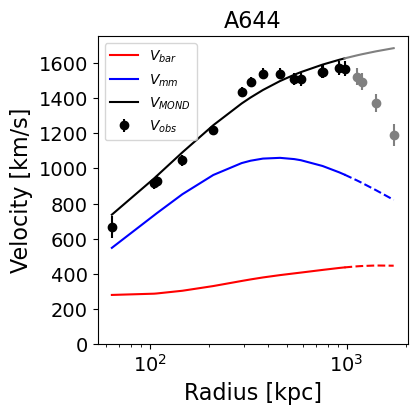}
    \includegraphics[width=4.5cm, height=4.5cm]{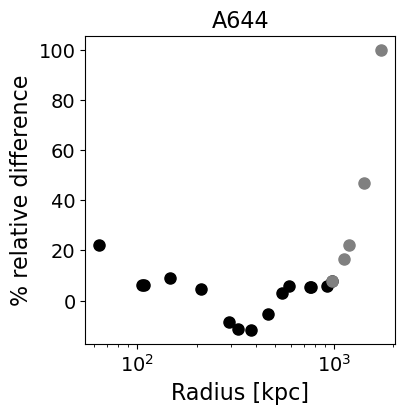}
    
	\includegraphics[width=4.5cm, height=4.5cm]{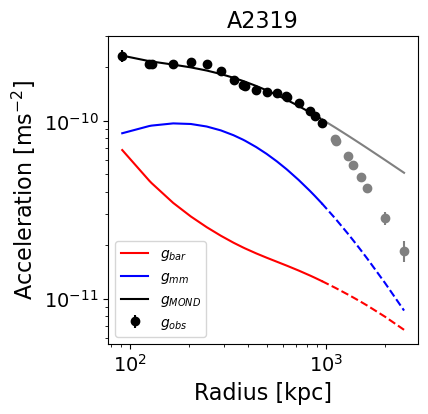}
	\includegraphics[width=4.5cm, height=4.5cm]{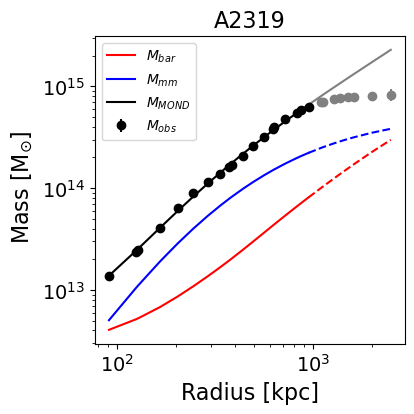}
	\includegraphics[width=4.5cm, height=4.5cm]{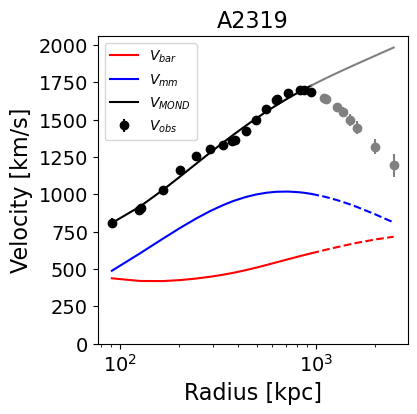}
	\includegraphics[width=4.5cm, height=4.5cm]{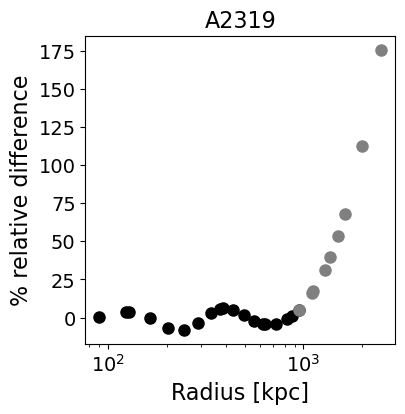}
          
	\caption{MCMC fit results for the isolated case (Eq.\,\ref{eq:nu}). From left to right: Acceleration profiles, enclosed mass profiles, circular-velocity profiles, and relative residual difference. In the first three panels, observed data (black points) are fitted with a mass model (black line) that includes the contribution of visible baryons (red line) and missing mass (blue line). We fit the data up to 1 Mpc (black points), where the hydrostatic bias should play a minor role, and then extrapolate the model to larger radii (grey line) to estimate the maximum hydrostatic bias in MOND (rightmost panels).}
\label{fig:fit_iso}
\end{figure*}

\begin{figure*}
\centering		 
	\includegraphics[width=4.5cm, height=4.5cm]{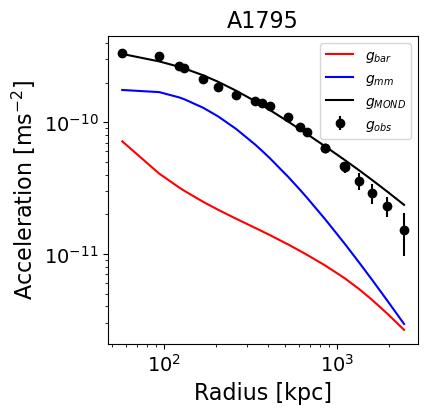}
	\includegraphics[width=4.5cm, height=4.5cm]{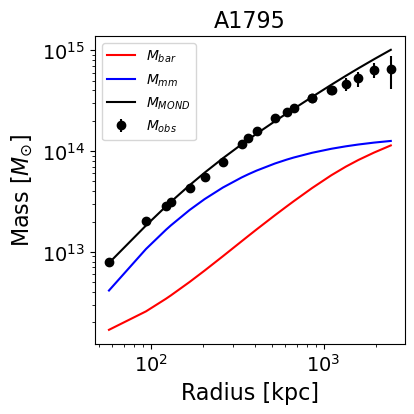}
	\includegraphics[width=4.5cm, height=4.5cm]{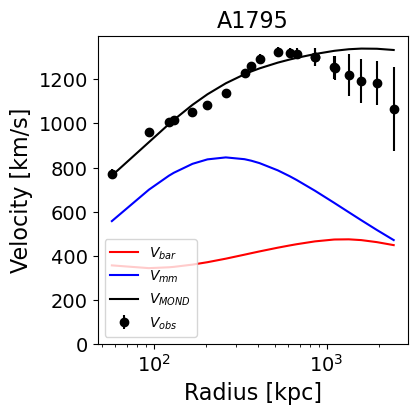}
	\includegraphics[width=4.5cm, height=4.5cm]{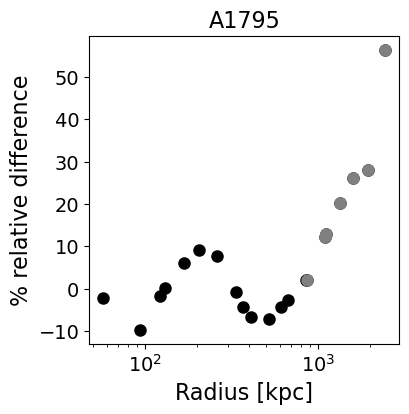}
          
	\includegraphics[width=4.5cm, height=4.5cm]{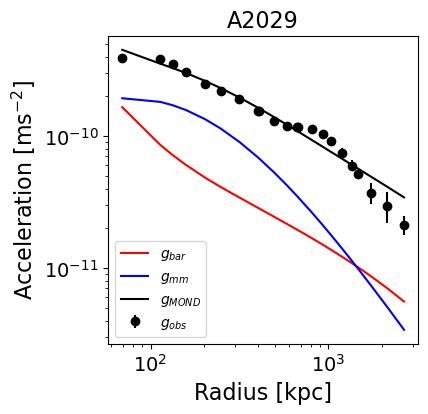}
	\includegraphics[width=4.5cm, height=4.5cm]{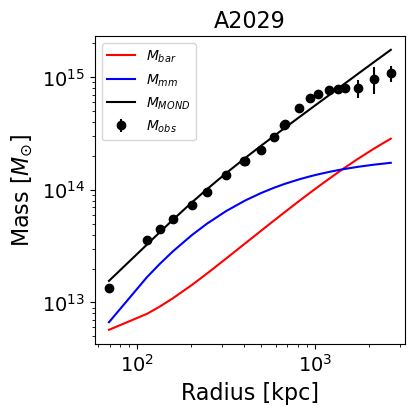}
	\includegraphics[width=4.5cm, height=4.5cm]{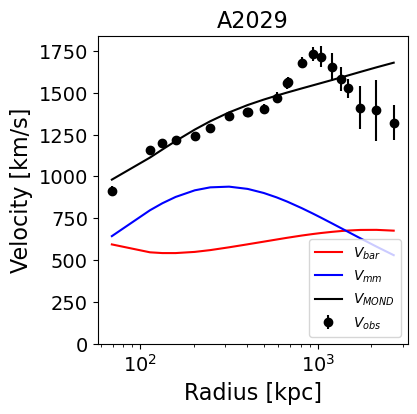}
	\includegraphics[width=4.5cm, height=4.5cm]{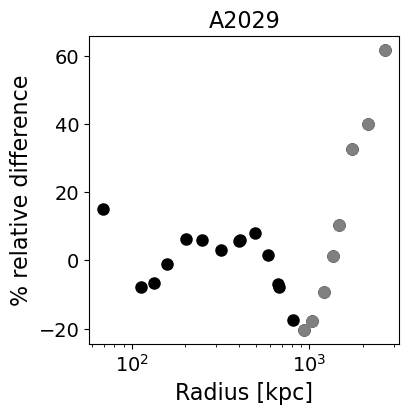}
          
	\includegraphics[width=4.5cm, height=4.5cm]{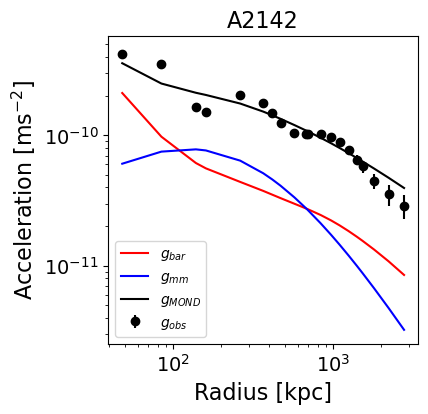}
	\includegraphics[width=4.5cm, height=4.5cm]{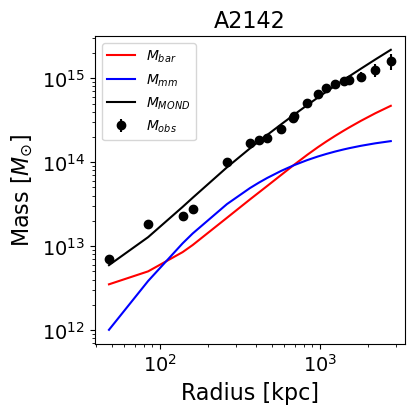}
	\includegraphics[width=4.5cm, height=4.5cm]{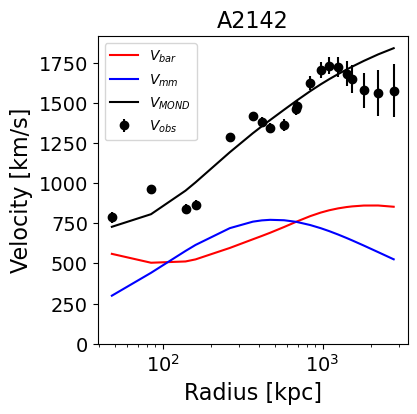}
	\includegraphics[width=4.5cm, height=4.5cm]{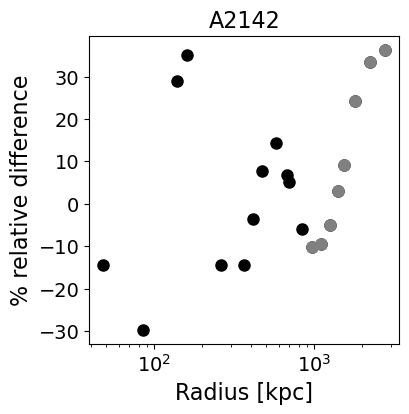}

	\includegraphics[width=4.5cm, height=4.5cm]{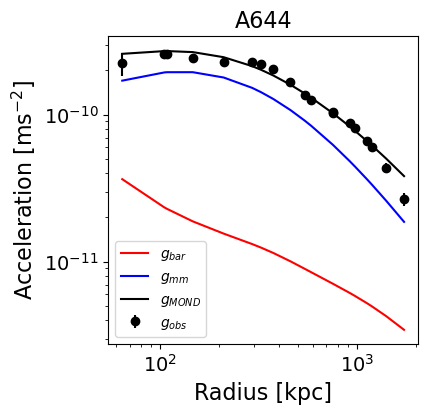}
	\includegraphics[width=4.5cm, height=4.5cm]{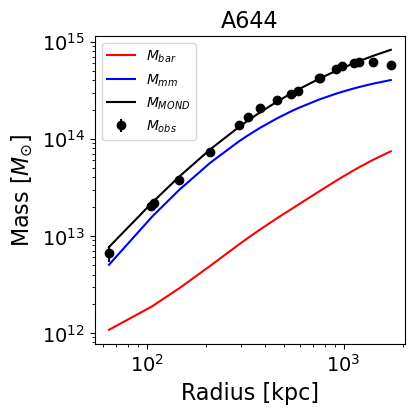}
	\includegraphics[width=4.5cm, height=4.5cm]{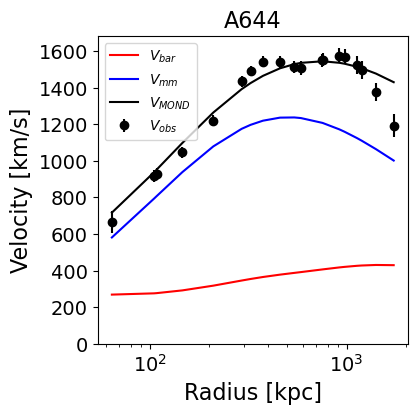}
    \includegraphics[width=4.5cm, height=4.5cm]{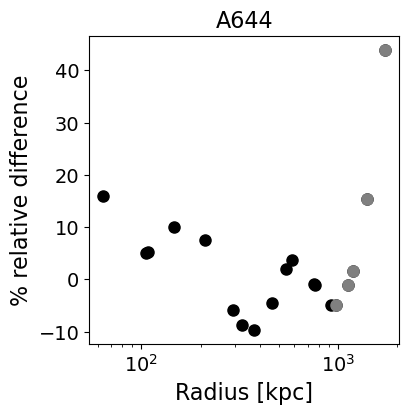}
    
	\includegraphics[width=4.5cm, height=4.5cm]{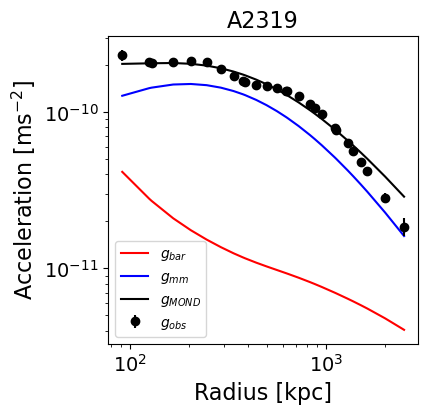}
	\includegraphics[width=4.5cm, height=4.5cm]{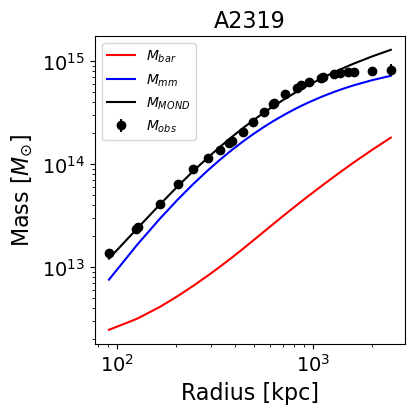}
	\includegraphics[width=4.5cm, height=4.5cm]{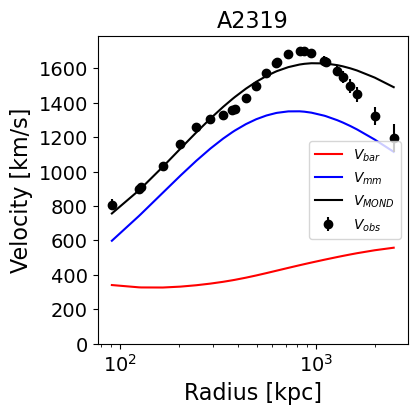}
	\includegraphics[width=4.5cm, height=4.5cm]{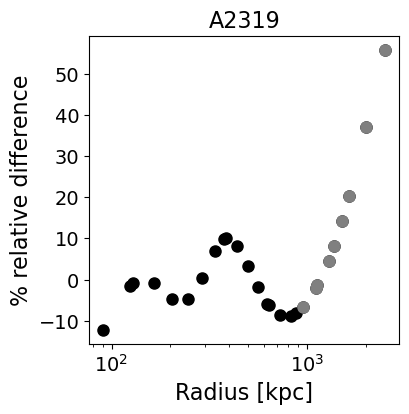}
          
	\caption{Same as Fig.\,\ref{fig:fit_iso} but for the non-isolated case with a basic modelling of the EFE (Eq.\,\ref{eq:nuEFE}). In this case we fit all data points, even those at $r>1$ Mpc, because we aim to estimate the minimum amonut of hydrostatic bias in MOND (rightmost panel).}
\label{fig:fit_efe}
\end{figure*}
\section{Results}\label{sec:results}

\subsection{MOND fits with no EFE}\label{sec:MCMCfits}

Figure \,\ref{fig:fit_iso} shows the MCMC results for the isolated, no-EFE case (Eq.\,\ref{eq:nu}). The fits are performed in acceleration space, but we also show the corresponding enclosed mass and circular velocity profiles from Eq.\,\ref{eq:shelltheorem}. We stress that $M_{\rm obs}$ and $M_{\rm MOND}$ are fictitious dynamical concepts: the former represents the Newtonian mass inferred from the observed acceleration (Eq.\,\ref{eq:hydro}) while the latter is that inferred from the MOND-predicted acceleration. In fact, both $M_{\rm obs}$ and $M_{\rm MOND}$ do not correspond to the sum of $M_{\rm bar}$ and $M_{\rm mm}$ (the observable masses of the physical components) because of the MOND boost. On the other hand, $V_{\rm obs}$ and $V_{\rm MOND}$ represent the `measurable' velocities that a test particle on a circular orbit should display to be in equilibrium with the cluster gravitational field. 

In general, the observations are well-fitted at $r<1$ Mpc. The rightmost panels in Fig.\,\ref{fig:fit_iso} show the relative residuals in percentage, given by
\begin{equation}
\delta =  100 \cdot  \frac{g_{\rm M} - g_{\rm obs}}{g_{\rm obs}}.
\end{equation}
Considering mass residuals rather than acceleration residuals would give the same results (cf. with Eq.\,\ref{eq:shelltheorem}). In general, the residuals at $r<1$ Mpc are within $\pm$20\%, demonstrating that the fits are acceptable considering usual astronomical accuracy. In fact, acceleration residuals of about 10-20$\%$ may be due to mere deviations from spherical symmetry rather than hydrostatic bias. The residuals at $r>1$ Mpc, where the MOND fit is extrapolated, quantify the maximum amount of hydrostatic bias in the case of no EFE. Such a maximum hydrostatic bias ranges from about 10\% to 100\%, apart for A2319 that is known to be involved in a major merger (see Sect.\,\ref{sec:sample}).

In some clusters (A2029 and A2142), the observed profiles show `bumps and wiggles' that may not necessarily trace the equilibrium gravitational potential (i.e. a static high-density mass shell) but may rather be driven by local deviations from the hydrostatic equilibrium or some other idiosyncrasies in the observations. Our best-fit models generally pass through those bumps and wiggles, so the final parameters should not be heavily affected by them.

\subsection{MOND fits with EFE}

Figure \,\ref{fig:fit_efe} shows the MCMC results for the EFE case (Eq.\,\ref{eq:nuEFE}). In the EFE case we consider all data, even those at $r>1$ Mpc, because we aim to quantify the minimum amount of hydrostatic bias needed in MOND when a maximal external field is taken into account. The observations are well fitted at all radii apart from the outermost ones. In the EFE case, the hydrostatic bias generally ranges from 10\% to 40\% up to $R\simeq2$ Mpc. This level of hydrostatic bias is sensible and comparable to that inferred in a $\Lambda$CDM context, as we discuss in Sect.\,\ref{sec:hydrobias}. For the very outermost points beyond 2 Mpc, the hydrostatic bias may go up to 50-60\% but these points are more uncertain due to the lower S/N ratio of the SZ effect.

For well-behaved clusters without known merging signatures (A1795, A2029, and A2142), the values of $g_{\rm N, e}$ are around $0.001-0.002 a_0$ and consistent with those inferred from the large-scale distribution of baryonic mass in the nearby Universe \citep[see Figures 7 and 8 in][]{Chae2021}, as well as those inferred from galaxy rotation-curve fits \citep{Chae2021, Chae2022}. For clusters with known merger signatures (A644, A2319), the values of $g_{\rm N, e}$ increase to $0.1-0.3 a_0$. These high values may indicate either a real EFE signal due to the merging subcluster, or be a `compensating' fitting artifact due to out-of-equilibrium dynamics. In any case, it is remarkable that known merging clusters clearly stand out in the sample in terms of $g_{\rm N, e}$, suggesting that this quantity is not merely an additional free parameter but could rather carry proper physical information.

\begin{figure}
\centering
	\includegraphics[width=0.23\textwidth]{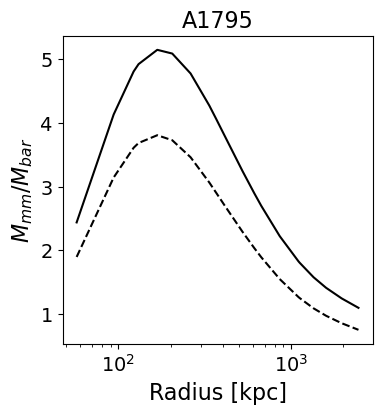}\\
	\includegraphics[width=0.23\textwidth]{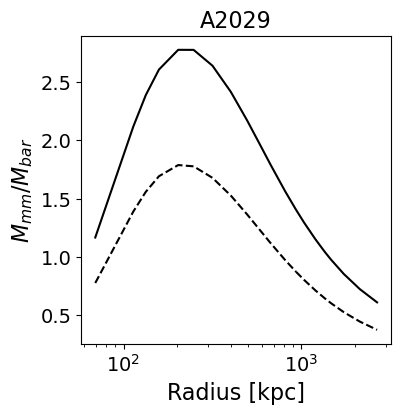}
	\includegraphics[width=0.23\textwidth]{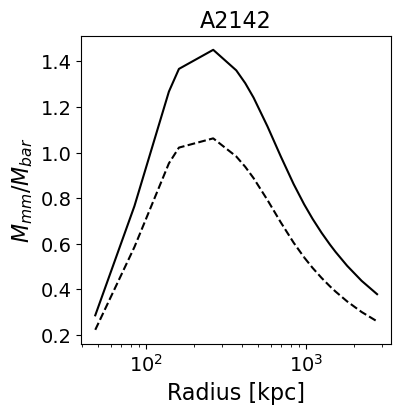}\\
	\includegraphics[width=0.23\textwidth]{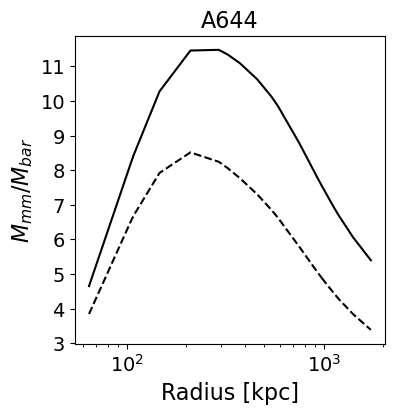}
 \includegraphics[width=0.23\textwidth]{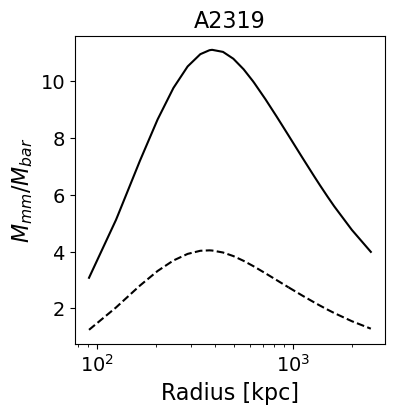}
 \caption{Missing-to-baryonic mass ratio in galaxy clusters as a function of radius in the EFE case (solid line) and isolated case (dashed line). The high values of A644 and A2319 should be interpreted with caution because these two galaxy clusters display merger signatures.}
\label{fig:mass_ratio}
\end{figure}

The ratio of missing-to-baryonic mass ($M_{\rm mm}/M_{\rm bar}$) as a function of radius is shown in Fig.\,\ref{fig:mass_ratio}. The shape of these profiles explicitly show that the missing matter is more centrally concentrated than the ICM, as suggested by the early work of \citet{Sanders1999, Sanders2003}. When the EFE is taken into account, the values of $M_{\rm mm}/M_{\rm bar}$ are systematically higher than for the isolated case because the MOND boost is decreased, so more missing mass is required (see also Fig.\,\ref{fig:EFE}). For non-merging clusters (A1795, A2029, and A2142), $M_{\rm mm}/M_{\rm bar}$ reaches a peak value of about $1-5$ at $r\simeq200-300$ kpc and decreases to about $0.4-1.5$ at $r\gtrsim 1$ Mpc, indicating that the total amount of missing mass is comparable to or smaller than the ICM mass. For merging clusters (A644 and A2319), the ratio $M_{\rm mm}/M_{\rm bar}$ display systematically higher values at all radii, which may again be driven by non-equilibrium dynamics rather than by actual missing mass. In any case, it is remarkable that known galaxy mergers (again) stand out in terms of MOND-inferred physical properties.

\begin{figure*}
\centering		 
	\includegraphics[width=0.32\textwidth]{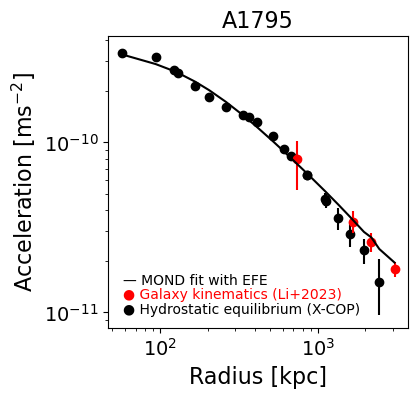}
	\includegraphics[width=0.32\textwidth]{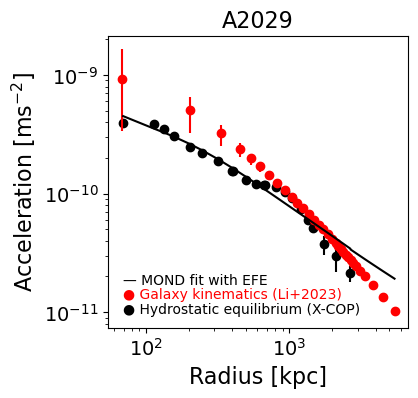}
	\includegraphics[width=0.32\textwidth]{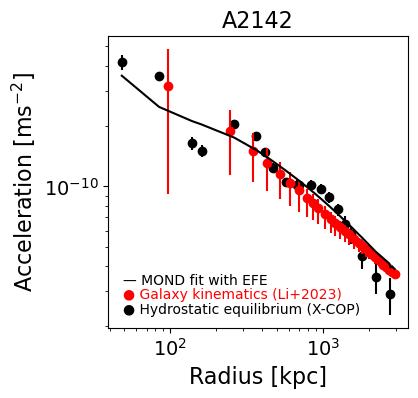}
	\caption{Comparison between acceleration profiles from hydrostatic equilibrium (black dots; from X-COP) and Jeans modelling of galaxy kinematics (red dots; from \citealt{Li2023}). The black line is the MOND EFE model fitted to the X-COP data (same as in Fig.\,\ref{fig:fit_efe}). Remarkably, for A1795 and A2142, the MOND model at large radii predicts the behaviour of the stellar kinematic data. For A2029 the comparison between the two datasets is poor, probably due to incompleteness in the galaxy kinematic data and/or unusual galaxy velocity anisotropy \citep[see][]{Li2023}.}
\label{fig:galaxykinematics}
\end{figure*}
\subsection{Comparison with galaxy kinematics data}

Another approach to measure the total mass profiles of galaxy clusters (so their acceleration profiles) is to model the observed kinematics of its galaxy members. Recently, \citet{Li2023} studied a sample of 16 galaxy clusters from HIFLUGCS \citep{Tian2021}, which includes the non-merging clusters in our sample (A1795, A2029, and A2142). In particular, \citet{Li2023} solved the spherical Jeans equations parametrising the galaxy velocity anisotropy and using two `virial shape parameters', which ameliorate the well-known mass-anisotropy degeneracy.

Figure\,\ref{fig:galaxykinematics} compares the acceleration profiles from X-COP using hydrostatic equilibrium with those from \citet{Li2023} using galaxy kinematics. For A1795 and A2142, the two methods generally agree, but at large radii ($r\gtrsim 1$ Mpc) galaxy kinematics give systematically higher accelerations than the hydrostatic equilibrium, pointing to hydrostatic bias (see also Sect.\,\ref{sec:hydrobias}). Remarkably, the MOND model reproduces the galaxy kinematics data better than the X-COP data at large radii, despite it was fitted to the latter dataset, not the former one. This fact testifies the predictive power of MOND even on galaxy cluster scales.

For A2029, the two methods significantly disagree. We cannot tell which one of the two profiles (if any) is the most reliable. However, we note that A2029 has been extensively discussed in \citet{Li2023} because it clearly stands out from the rest of their sample. Firstly, A2029 is the only galaxy cluster in \citet{Li2023} for which radially varying incompleteness may be a concern (see their Appendix A). Secondly, A2029 is the only galaxy clusters in \citet{Li2023} that show a strong radial variation in the galaxy velocity anisotropy, going from $-0.4$ at small radii to $0.7$ at large radii (see their Figure 4). This behaviour in anisotropy is quite unusual and cast some doubts on the enclosed mass profile, given the mass-anisotropy degeneracy. Future investigations of A2029 using both hydrostatic equilibrium and galaxy kinematics may share new light on this cluster.

\section{Discussion}\label{sec:disc}

\begin{figure}
\centering
	\includegraphics[width=0.45\textwidth]{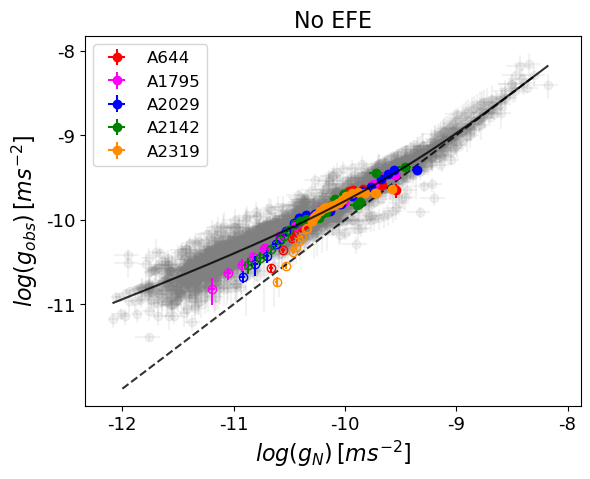}
	\includegraphics[width=0.45\textwidth]{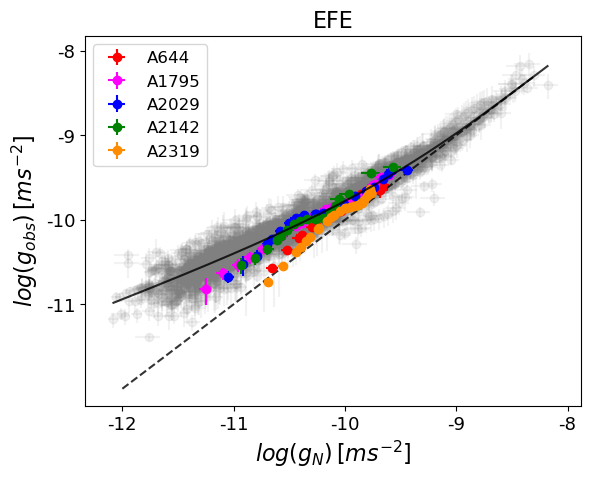}
\caption{Locations of galaxy clusters on the radial acceleration relation after accounting for a missing mass component. The top panel corresponds to the isolated case (Eq.\,\ref{eq:nu}), while the bottom panel to the EFE case (Eq.\,\ref{eq:nuEFE}). Grey points show disc galaxies from the SPARC database \citep{Lelli2016}, while coloured points show the five galaxy clusters in our sample. The solid line is the MOND prediction for isolated systems; the dashed line is the line of unity (Newtonian prediction with no DM). In the top panel, open symbols indicate data at $R>1$ Mpc, where the mass models with no EFE have been extrapolated.}
\label{fig:RAR}
\end{figure}

It has been known for decades that MOND can greatly reduce the need for DM in galaxy clusters but cannot entirely eradicate it. While the global mass discrepancies ($M_{\rm tot}/M_{\rm bar}$) are typically about 5-10 in Newtonian dynamics, they are reduced to a factor of about 2 in MOND \citep{Sanders1999}. In this work, we studied the spatial distribution of the missing matter required in MOND. Physically acceptable solutions for the missing mass profile can be trivially found as long as there is some modest level of hydrostatic bias at large radii ($r>1$ Mpc). The situation is summarised in Fig.\,\ref{fig:RAR}, which shows the location of our galaxy clusters on the RAR defined by disc galaxies \citep{Lelli2017}. After a sensible missing mass component is included, galaxy clusters lie on the RAR within the observed scatter. At low acceleration (large radii), the cluster data systematically deviate below the average RAR, which may be due to hydrostatic bias and/or the EFE \citep[e.g.][]{Chae2020, Chae2021}.

If the missing mass required by MOND is a real physical entity (rather than a more complex modification of the gravitational law), then its nature needs to be identified. In the following, we discuss the viability and the existing constraints on (1) undetected `missing baryons' (Sect.\,\ref{sec:darkbaryons}), (2) standard active neutrinos (Sect.\,\ref{sec:neutrinos}), and (3) sterile neutrinos (Sect.\,\ref{sec:sterile}). Finally, we discuss the long-standing issue of hydrostatic bias in both $\Lambda$CDM and MOND contexts (Sect.\,\ref{sec:hydrobias}).

\subsection{The nature of the missing mass: Undetected baryons}\label{sec:darkbaryons}

The $\Lambda$CDM cosmological model has a well-studied `missing baryons' problem: the total amount of baryons observed in galaxies, galaxy groups, and galaxy clusters makes up only $\sim$18$\%$ of the cosmic baryon density ($\Omega_{\rm bar}$) expected from $\Lambda$CDM fits to the CMB and from big bang nucleosynthesis (BBN) calculations \citep[e.g.][]{Fukugita2004}. The standard solution is that the vast majority of baryons reside in the diffuse warm-hot intergalactic medium (WHIM) that exist in between galaxies and galaxy clusters \citep{Shull2012, Macquart2020}, but whose precise amount is still subject to significant uncertainties.

In a MOND cosmology, one may expect a similar missing baryon problem because the physics of BBN primarily depends on the baryon-to-photon ratio $\Omega_{\rm bar}/\Omega_{\gamma}$ and to a minor degree on the details of the expansion history, so on the underlying gravitational law\footnote{More generally, any cosmological model with standard BBN and the usual interpretation of the CMB radiation (providing $\Omega_{\rm \gamma}$) is expected to display a missing baryons problem.}. In this case, the amount of missing baryons that one would need inside galaxy clusters is of the order of only 1-4$\%$ $\Omega_{\rm bar}$, representing a small fraction of the cosmologically available missing baryons (82$\%$ $\Omega_{\rm bar}$). In such a MOND cosmology, the bulk of the baryons should still reside in the WHIM.

\citet{Milgrom2008} proposed that the missing baryons in galaxy clusters could be made of dense clouds of cold gas. A multi-phase, multi-temperature ICM is a sensible possibility to consider. For example, the intergalactic medium (IGM) around isolated galaxies is known to be multi-phase, showing evidence for cold gas clouds with $T\simeq10^4$ K embedded in a hot diffuse plasma with $T\simeq10^{6}-10^{7}$ K \citep[e.g.][]{Afruni2019, Afruni2021, Afruni2022}. State-of-the-art \hi\ surveys of the nearest galaxy clusters reach a point-source mass limit of $0.5-1.0\times10^{6}$ M$_\odot$ at distances of $15-20$ Mpc \citep{Kleiner2023, Boselli2023}. Then, if the gas clouds proposed by \citet{Milgrom2008} are made of atomic hydrogen, they must be less massive than $5\times10^{5}$ M$_\odot$. The upcoming square kilometre array (SKA) will improve the \hi\ mass limits by at least one order of magnitude \citep[e.g.][]{Blyth2015}. If the cold gas clouds are made of molecular hydrogen (H$_2$), instead, they would be more difficult to detect because they may share the same metallicity of the ICM ($\sim$0.3 $Z_\sun$), so their CO emission would be weak due to variations in the CO-to-H$_2$ conversion factor \citep[e.g.][]{Bolatto2013}. In addition, there should be a mechanism that prevents the molecular clouds to collapse and form stars (otherwise they would emit UV radiation), so \hi\ clouds may be more appealing than H$_2$ ones.

\citet{Milgrom2008} noted that the kinetic energy of these hypothetical clouds is about ten times larger than the thermal energy of the ICM within 200-300 kpc, providing a substantial energy reservoir. Then, if the gas clouds interact at a sufficient rate (via cloud-cloud collisions or cloud-ICM dynamical friction), their kinetic energy can be converted into thermal energy, heating up the ICM and solving the long-standing `cooling flow' problem in galaxy clusters \citep{Fabian1994}. Interestingly, this type of heating source would be smoothly distributed in the cluster core and steady with time, in contrast to feedback from active galactic nuclei that is generally proposed to solve the cooling flow problem in a $\Lambda$CDM context \citep{Sijacki2007}. Assuming that the kinetic heating rate from collisions balances the cooling rate of the ICM, \citet{Milgrom2008} finds that the area covering factor of the clouds $f_{\rm A}$ is about $10^{-3}-10^{-4}$ depending on the ratio $M_{\rm mm}/M_{\rm bar}$ in the cluster core. The volume filling factor $f_{\rm V}$ of the clouds must be even smaller than this because $f_{\rm V}\simeq f_{\rm A} r_{\rm cl}/(2 r_{\rm s})$, where $r_{\rm c}$ is the typical cloud radius and $r_{\rm s}$ is the scale radius of the missing mass component (Table\,\ref{tab:fit}), so $r_{\rm cl}/(2 r_{\rm s})\ll1$. Intriguingly, cold gas clouds with similarly low volume filling factors have been inferred to exist inside the giant Ly$\alpha$-emitting nebulae of ionised gas surrounding massive quasar-host galaxies at $z\simeq3$ \citep{Pezzulli2019}, which are presumably the progenitors of local BCGs at the center of galaxy clusters.

We propose an additional argument to infer the cloud properties: assuming that they are pressure confined by the hot ICM in the same way as the \hi\ high-velocity clouds around the Milky Way are thought to be pressure confined by its hot corona \citep[e.g.][]{Spitzer1956}. Pressure equilibrium therefore implies that
\begin{equation}
\dfrac{n_{\rm cold}}{n_{\rm hot}} = \dfrac{T_{\rm hot}}{T_{\rm cold}},
\end{equation}
where $T_{\rm cold}$ and $n_{\rm cold}$ are the unknown temperatures and hydrogen number density of the cold gas clouds, while $T_{\rm hot}\simeq10^8$ K and $n_{\rm hot}\simeq10^{-3}$ cm$^{-3}$ are typical values for X-ray-emitting gas. If $T\simeq10^3-10^4$ K as for typical \hi\ gas, then $n_{\rm cold}\simeq10-100$ cm$^{-3}$. If $T\simeq10-100$ K as for typical H$_2$ gas, then $n_{\rm cold}\simeq10^3-10^4$ cm$^{-3}$. If we consider $T\simeq10^4$ K and a conservative \hi\ upper limit of $M_{\rm c} < 10^5$ M$_{\odot}$, we expect a cloud radius $R_{\rm c} < 50$ pc. Clouds with lower masses and/or lower temperatures (such as molecular clouds) would have even smaller radii. These small sizes are consistent with the low values of $f_{\rm A}$ inferred by \citet{Milgrom2008} and make these clouds very difficult to detect in both emission (because they would not properly `fill the beam' of existing radio/mm interferometers at cluster distances) and absorption (because they would have a very small chance to align with a bright background source such as a quasar).

\begin{table*}
\centering
\caption{Results of correlation tests between the 3D temperature of the ICM and the properties of the missing mass component (see Fig. 5)}
\begin{tabular}{l|cc|cc|cc|cc}
Correlation & \multicolumn{2}{c|}{Pearson} & \multicolumn{2}{c|}{Spearman} & \multicolumn{2}{c|}{Kendall} & \multicolumn{2}{c}{Fit Results: $y=s x + N$}\\
& $\rho_{\rm P}$ & $p$-value & $\rho_{\rm S}$ & $p$-value & $\tau_{\rm K}$ & $p$-value & s & $N$ \\
\hline
$\log(T_{\rm 3D})-\log(M_{\rm mm})$ with EFE & 0.73 & 0.16 & 0.70 & 0.19 & 0.60 & 0.23 &  3.93 &-51.78\\
$\log(T_{\rm 3D})-\log(M_{\rm mm})$ no EFE & 0.73 & 0.16 & 0.70 & 0.19 & 0.60 & 0.23 & 5.38& -72.01\\
$\log(T_{\rm 3D})-\log(r_{\rm s})$ with EFE & 0.86 & 0.063 & 0.90 & 0.037 & 0.80 & 0.083 & 7.23 &-11.42\\
$\log(T_{\rm 3D})-\log(r_{\rm s})$ no EFE & 0.86 & 0.063 & 0.90 & 0.037 & 0.80 & 0.083 & 7.85 & -12.60\\
\end{tabular}
\label{tab:corrtest}
\end{table*}
\begin{figure*}
\centering
	\includegraphics[width=4.5cm, height=4.5cm]{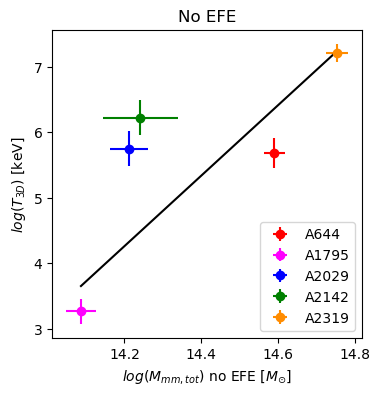}
	\includegraphics[width=4.5cm, height=4.5cm]{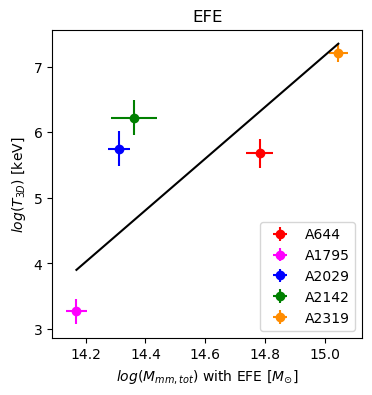}
	\includegraphics[width=4.5cm, height=4.5cm]{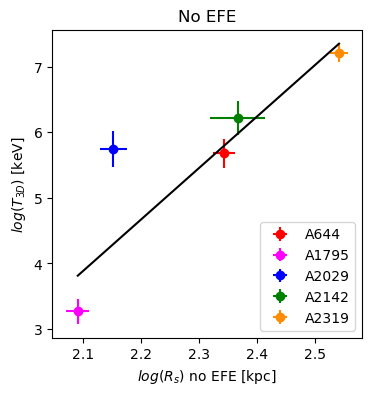}
    \includegraphics[width=4.5cm, height=4.5cm]{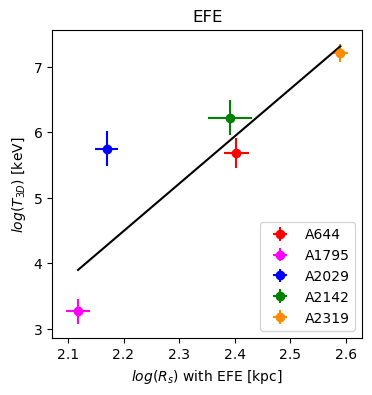}
    \label{fig:corr}
\caption{Relations between the 3D temperature of the ICM and the properties of the missing mass component: Total mass $M_{\rm mm}$ (leftmost panels) and scale radius $r_{\rm s}$ (rightmost panels) for both the isolated and EFE case. In all cases there are statistically significant correlations but we warn that they are based only on five clusters, two of which (A644 and A2319) may be out of dynamical equilibrium.}
\label{fig:corr}
\end{figure*}
To explore the possible connection between missing mass and hot gas, we plotted $M_{\rm mm, tot}$ and $r_{\rm s}$ versus several observed ICM properties. We found a tentative correlation with the median value of the 3D temperature $T_{\rm 3D}$ (see Fig.\,\ref{fig:corr}). Formally, these relations are statistically significant, having high values of the Pearson's, Spearman's and Kendall's correlation parameters (given in Table\,\ref{tab:corrtest}). However, we stress that these values are based on only five galaxy clusters, two of which may be out of dynamical equilibrium, possibly providing unrealiable values for $M_{\rm mm}$ and $r_{\rm s}$. A similar $M_{\rm mm}-T_{\rm 3D}$ relation was reported by \citet{Angus2008} for a larger cluster sample, but using a different methodology and definitions of $M_{\rm mm}$ and $T_{\rm 3D}$. If these correlations are confirmed by future studies, they could point to a real connection between the properties of the missing matter and those of the ICM, possibly favouring the missing baryons hypothesis.

\subsection{The nature of the missing mass: Active neutrinos}\label{sec:neutrinos}

Neutrino oscillation experiments provide clear evidence for neutrino masses and flavour mixing. These experiments do not measure the actual neutrino masses $m_{\nu}$, but indicate that the largest mass difference between active neutrinos is $\Delta m_{\nu}\simeq0.05$ eV \citep[e.g.][]{GonzalezGarcia2003}. Since the number density of neutrinos produced in the early Universe is expected to be similar to that of photons, there must be a cosmic neutrino fluid with 
\begin{equation}\label{eq:Omega_nu}
 \Omega_{\rm \nu}h^2 = \dfrac{1}{94} \sum_i^{N_\nu} \dfrac{m_{\nu, \rm i}}{\rm eV}   
\end{equation}
where the sum is over $N_\nu$ neutrino types \citep{Sanders2003}. If $m_{\nu}\simeq \Delta m_\nu$, then $\Omega_{\nu}\simeq 10^{-3}$ and neutrinos have no significant cosmological mass density, so they cannot contribute to the mass of any bound system, as one expects in the $\Lambda$CDM context. The Planck data combined with baryonic acustic oscillations, indeed, imply that $\sum_i m_{\nu,i}< 0.12$ eV in a flat $\Lambda$CDM cosmology \citep{Planck2020} otherwise $\Omega_\nu$ would be too large and leave too little room to $\Omega_{\rm CDM}$, reducing the power on small scales. In a MOND context, these constraints on $\sum_i m_{\nu,i}$ do not apply, so one can consider the case where $m_\nu \gg \Delta m_\nu$ and the masses of the three active neutrinos are nearly equal. Then, cosmic neutrinos would behave like hot dark matter (HDM), undergoing gravitational instability and collapse on spatial scales that depend on their mass \citep[e.g.][]{Sanders2003}.

\citet{Sanders2003, Sanders2007} found that standard active neutrinos with $m_\nu\simeq2$ eV could provide the missing matter required by MOND on galaxy cluster scales ($R\simeq1$ Mpc), while they would be too light (too hot) to provide significant gravitational contributions on galaxy scales ($R\simeq100$ kpc), preserving the MOND successes in that regime. On the contrary, \citet{Takahashi2007}, \citet{Angus2008} and \citet{Natarajan2008} found that active neutrinos with $m_\nu\simeq2$ eV cannot entirely explain the MOND missing mass in the center of galaxy clusters, especially in low-mass and low-temperature ones. In any case, the latest results of the KATRIN experiment provide an upper limit of $m_\nu < 0.8$ eV at 90\% confidence level for the electron anti-neutrino mass \citep{KATRIN1}, ruling out active neutrinos with $m_\nu\simeq2$ eV. Hereafter, we briefly recap the basic neutrino argument and revisit the issue in light of our new measurements.

Cosmological neutrinos are created with a maximum phase-space density of $(2\pi\hbar)^{-3}$ per type (including antineutrinos) which is conserved during their subsequent evolution \citep{TremaineGunn1979}. During the formation of a collapsed object out of the mixed neutrino-baryon fluid, the two fluids are expected to attain the same velocity dispersion (temperature) via violent relaxation, so one can derive a relation between the maximum final density of the neutrino component and the velocity dispersion (temperature) of the system \citep[cf.][]{Sanders2003, Sanders2007}:
\begin{equation}\label{eq:TremaineGunnLimit}
    \dfrac{\rho_{\nu}}{\rm g\,cm^{-3}} \leq 10^{-28} \left( \dfrac{T}{{\rm keV}} \right)^{1.5} \sum_i^{N_\nu} \left(\dfrac{m_{\nu, \rm i}}{\rm eV}\right)^4,
\end{equation}
or equivalently
\begin{equation}
    \dfrac{\rho_{\nu}}{\rm M_\odot\,pc^{-3}} \leq (1.5\times10^{-6}) \left( \dfrac{T}{{\rm keV}} \right)^{1.5} \sum_i^{N_\nu} \left(\dfrac{m_{\nu, \rm i}}{\rm eV}\right)^4.
\end{equation}
If we consider three active neutrinos with the same mass $m_{\nu, {\rm a}}$, the mean $T_{\rm 3D}$ of the X-COP galaxy clusters (Table\,\ref{tab:properties}), and our measurements of $\rho_0$ (Table\,\ref{tab:fit}), we find $m_{\nu, {\rm a}}>2-5$ eV that is surely ruled out by the KATRIN upper limit. Thus, it is clear that active neutrinos cannot form the missing mass needed by MOND in galaxy clusters, confirming the findings of \citet{Takahashi2007}, \citet{Angus2008} and \citet{Natarajan2008}.

\subsection{The nature of the missing mass: Sterile neutrinos}\label{sec:sterile}

Another proposal is that of sterile neutrinos \citep{Angus2008}: hypothetical right-handed Fermions that are neutral under both weak interactions (`sterile') and electromagnetic ones (`neutrinos'). Sterile neutrinos are empirically motivated by observed anomalies in neutrino oscillations and can be trivially added to the standard model of particle physics \citep[e.g.][]{Boser2020, Basudeb2021}. Assuming that MOND exactly converges to General Relativity (GR) in the early Universe, \citet{Angus2009} showed that a single type of sterile neutrino can fully replace CDM in fitting the CMB data from WMAP, providing $\Omega_{\nu_s}h^2=0.117$ and $m_{\nu, \rm s}\simeq11$ eV (in the case of thermal production for which Eq.\,\ref{eq:Omega_nu} holds). The CMB fit of \citet{Angus2009} fostered the investigation of a MOND cosmology supplemented by 11-eV sterile neutrinos, dubbed $\nu$HDM, which lead to some promising results but also severe challenges \citep{Angus2011, Katz2013, Angus2014, Haslbauer2020, Asencio2021, Wittenburg2023}.

Importantly, the most recent CMB data from Planck strongly constrain the equation of state of a generic  `dark fluid' during recombination, showing no significant deviation from a dust fluid (CDM) and leaving little room for HDM \citep{Thomas2016, Kopp2018, Ilic2021}. In addition, relativistic extensions of MOND may significantly deviate from GR in the early Universe \citep{relativisticmond}, so the constraints on neutrino masses (active and sterile) are expected to be different and need to be recomputed accounting for CMB lensing.

Our measurements of $\rho_0$ and the Gunn-Tremaine limit (Eq.\,\ref{eq:TremaineGunnLimit}) imply that $m_{\nu, {\rm s}} > 10\pm3$ eV, considering the average mass from our five galaxy clusters and its standard deviation. This lower limit is consistent with $m_{\nu, {\rm s}}\simeq11$ eV. Similar results were found by \citet{Angus2010} using self-consistent models of semi-degenerate neutrinos in equilibrium with the clusters' gravitational potential. To date, the KATRIN experiment indicates that sterile neutrinos with $m_{\nu, \rm s}\simeq11$ eV remain a viable possibility only for small values of active-to-sterile neutrino mixing \citep{KATRIN2}, which actually is the regime where the basic argument of \citet{Angus2009} holds. Stronger constraints on sterile neutrinos will be available in future KATRIN runs, so we will know whether a mass of 11 eV is viable in the coming years.

\subsection{Hydrostatic bias: Insights from both MOND and $\Lambda$CDM}\label{sec:hydrobias}

The issue of hydrostatic bias has been amply studied in the standard cosmological context \citep[e.g.][]{Smith2016, Eckert2016, Henson2017, Angelinelli2020, Barnes2021}. Clusters' mass profiles are obtained under the assumption that the hot gas is in hydrostatic equilibrium with the gravitational potential, but this assumption may break down for a variety of reasons. For example, there may be non-thermal gravitational support that is unaccounted for by Equations \ref{eq:hydro} and \ref{eq:idealgas}. Such non-thermal support may be due to gas turbulence, bulk gas motions, or even gas rotation \citep{Bartalesi2023}. In addition, the ICM is not expected to be completely smooth and uniform, but to contain gas clumps that can alter the inferred gas density $\rho_{\rm gas}$ \citep[e.g.][]{Roncarelli2013, Eckert2015, Towler2023}. Indeed, the observed X-ray emissivity $\epsilon$ is proportional to $\rho_{\rm gas}^2$, so small variations in $\rho_{\rm gas}$ can lead to large variations in $\epsilon$. Finally, some galaxy clusters may be undergoing mergers in their outskirts, so they are not entirely relaxed structures in dynamical equilibrium, as it is probably the case for A644 and A2319 in our sample (see Sect.\,\ref{sec:sample}). These possible effects, among others, may cause the assumption of hydrostatic equilibrium to be invalid. Importantly, these effects are not unique to $\Lambda$CDM, nor to MOND, so they can play a role in both contexts.

In a $\Lambda$CDM context, hydrostatic bias is often invoked to solve two related cosmological problems: (1) discrepancies in the $\sigma_8$ parameter (the normalisation of the power spectrum of density fluctuations) obtained from cluster counts with respect to that obtained from CMB fits \citep{Planck2014}, and (2) discrepancies between the baryonic fraction measured in galaxy clusters (the ratio between observed baryonic mass and halo virial mass) with the cosmic baryonic fraction $\Omega_{\rm bar}/\Omega_{\rm m}$ from CMB fits \citep[e.g.][]{Eckert2019, Li2023, Wicker2023}. The amount of hydrostatic bias has been measured using a variety of methods, leading to contradictory results, but typically range between 10-50$\%$. In a MOND context, the level of hydrostatic bias may be different from that required in $\Lambda$CDM, but we actually find it to be comparable (10-40$\%$) with a clear radial dependence (Fig.\,\ref{fig:fit_efe}). 

In the specific case of the X-COP cluster sample, \citet{gravitationalxcop} investigated the issue of hydrostatic bias by comparing masses from hydrostatic equilibrium with those from weak lensing. They concluded that hydrostatic bias is important at radii beyond $R_{500}$, which is defined as the radius where the enclosed `dynamical mass' density is 500 times the critical density of the Universe. Such a radius approximately corresponds to 1 Mpc (see Table 2 in \citealt{gravitationalxcop}), similar to the radius at which our MOND fits imply an increasing level of hydrostatic bias. For A1795 and A2142, the effect of hydrostatic bias is confirmed by comparing the acceleration profiles from X-COP with those from galaxy kinematics (Fig.\,\ref{fig:galaxykinematics}), as also pointed out by \citet{Li2023}. For A2029, instead, the situation remains unclear. In any case, a sensible level of hydrostatic bias allows for a MOND missing mass component with a physical mass density profile that converges to a finite total mass, so the concerns raised in \citet{gravitationalxcop} might simply be due to hydrostatic bias.

\section{Conclusions} \label{sec:sum}

We built mass models of galaxy clusters in Milgromian dynamics (MOND). We focused on five clusters from the X-COP sample \citep{Ghirardini2018} for which both high-quality baryonic (stars in galaxies and hot gas in the ICM) and dynamical information (from hydrostatic equilibrium and the SZ effect) are available. This sample contains two merging clusters (A644 and A2319), which are studied only for the sake of comparison with clusters without known merger signatures (A1795, A2029, A2142). Our results can be summarised as follows:
\begin{enumerate}
    \item We confirm the well-known result that galaxy clusters require additional `missing matter' in MOND. Using a basic subtraction approach, the mass profiles of the missing matter decline at a rate of $r>1$ Mpc and are therefore unphysical. This effect largely vanishes considering the MOND EFE and/or sensible levels of hydrostatic bias at large radii.
    \item Using a Bayesian MCMC approach, we fit the acceleration profiles of the clusters, adding a missing mass component. We find good results using a density profile with an inner core and an outer slope of $-4$, which gives a finite total mass for the missing matter component. 
    \item MOND fits without the EFE imply a maximum amount of hydrostatic bias at $R>1$ Mpc of between 10$\%$$-$100$\%$, whereas MOND fits considering the EFE reduce the implied amount of hydrostatic bias at $R>1$ Mpc to 10$\%$$-$40$\%$. The required external field strengths ($g_{\rm Ne}/a_0\simeq10^{-3}$) are consistent with those expected from the large-scale baryonic mass distribution \citep{Chae2021} except for the two merging clusters, which may be out of dynamical equilibrium.
    \item For non-merging clusters, the missing-to-visible mass ratio ($M_{\rm mm}/M_{\rm bar}$) is about $1-5$ at $R\simeq200-300$ kpc and decreases to $0.4-1.1$ at large radii, indicating that the total amount of missing mass is smaller than or comparable to the ICM mass. For merging clusters, the values of $M_{\rm mm}/M_{\rm bar}$ are systematically higher but may be driven by out-of-equilibrium dynamics rather than actual missing mass.
\end{enumerate}
In conclusion, galaxy clusters do not seem to be an insurmountable challenge for MOND as long as there is a sensible extra component with similar mass to the hot gas. Such missing matter may be baryonic, such as pressure-confined dense clouds of cold gas, or may require a minimal extension of the standard model of particle physics, such as sterile neutrinos with $m_{\nu, {\rm s}}\gtrsim10$ eV. In comparison, in a $\Lambda$CDM cosmology, non-baryonic DM needs to be at least five times more abundant than the visible baryonic matter and cannot consist of light particles. We find tentative evidence for a possible correlation between the properties of the missing matter (mass and scale radius) and the temperature of the hot gas, suggesting that the missing matter may be a real physical entity related to the ICM properties. This tentative correlation could favour a purely baryonic interpretation of the missing matter, but we stress that it is based on only five galaxy clusters, two of which are possibly out of dynamical equilibrium. Clearly, a larger cluster sample with full baryonic information is needed to explore possible statistical correlations in a more comprehensive way and to shed new light on the nature of the missing mass in MOND.

\begin{acknowledgements}
We are grateful to Dominique Eckert and Stefano Ettori for providing the X-COP data, and to Pengfei Li for providing the galaxy kinematics data. We also thank Andrea Biviano, Harry Desmond, Stacy McGaugh, Moti Milgrom, James Schombert, Constantinos Skordis, and Paolo Tozzi for precious comments and suggestions about this work.
\end{acknowledgements}

\bibliographystyle{aa} 
\bibliography{references}

\begin{appendix}

\section{MOND missing mass profiles from a basic subtraction approach}\label{sec:sub}

In this section, we estimate the MOND missing mass profile using a basic subtraction approach, similarly to \citet{gravitationalxcop}. Differently from \citet{gravitationalxcop}, however, we use the non-parametric log-normal mixture reconstruction of $g_{\rm obs}(r)$ rather than the parametric Einasto reconstruction. In addition, we investigate the EFE that was neglected by \citet{gravitationalxcop}.

\begin{figure*}
\centering
	\includegraphics[width=4.5cm, height=4.5cm]{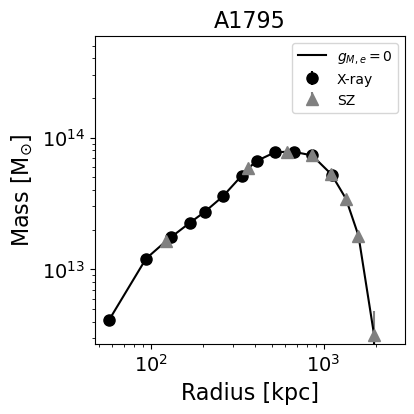}
    \includegraphics[width=4.5cm, height=4.5cm]{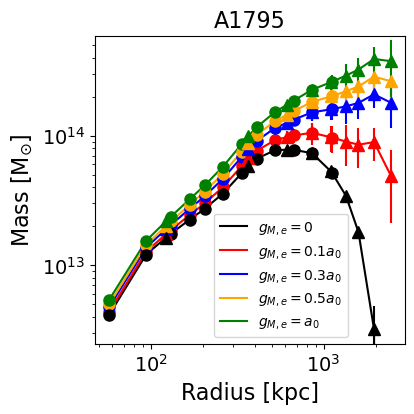}
		 
	\includegraphics[width=4.5cm, height=4.5cm]{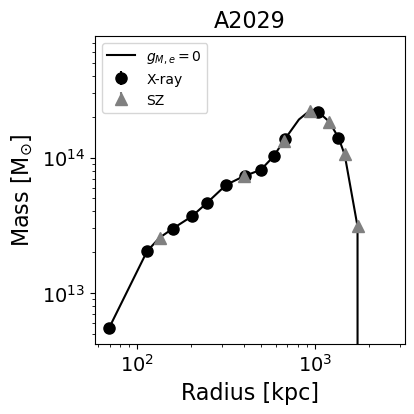}
	\includegraphics[width=4.5cm, height=4.5cm]{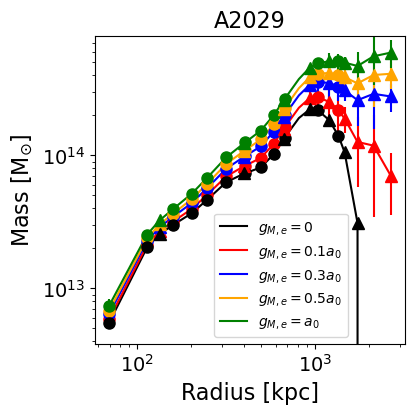}
	\includegraphics[width=4.5cm, height=4.5cm]{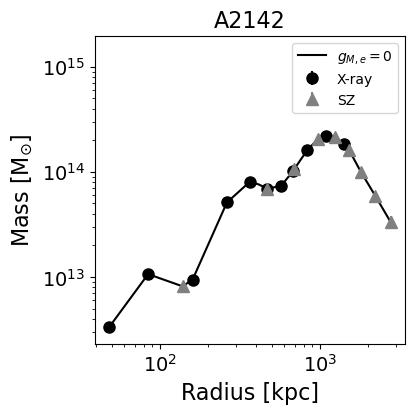}
	\includegraphics[width=4.5cm, height=4.5cm]{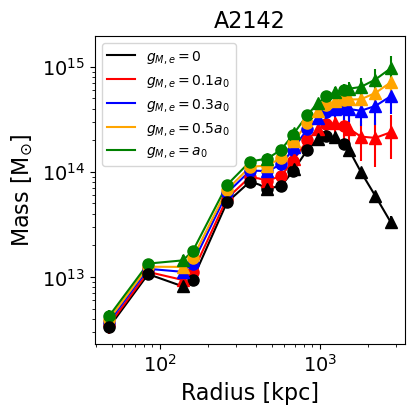}\\

    \includegraphics[width=4.5cm, height=4.5cm]{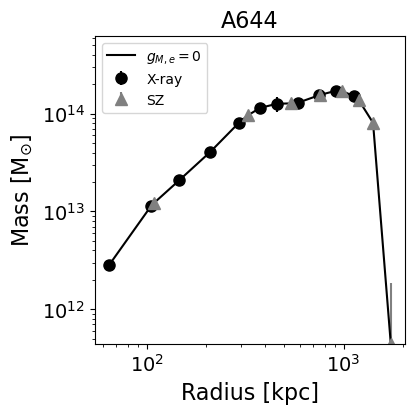}
	\includegraphics[width=4.5cm, height=4.5cm]{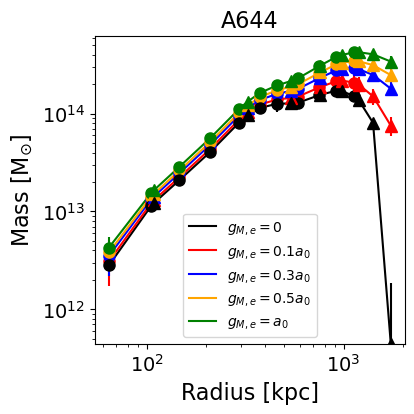}
	\includegraphics[width=4.5cm, height=4.5cm]{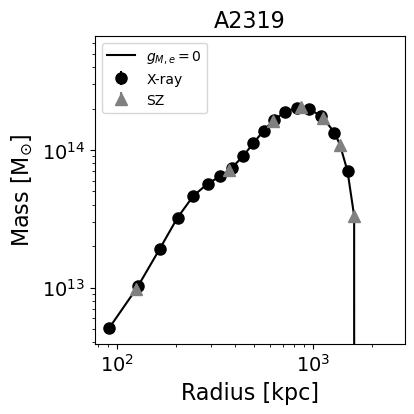}
	\includegraphics[width=4.5cm, height=4.5cm]{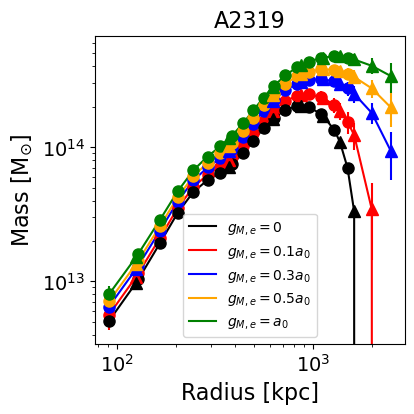}

	\caption{Enclosed mass profiles for the MOND missing matter using a simplistic subtraction approach (see Sect.\,\ref{sec:sub}). Dots show data from X‐ray observations only, while triangles show those adding the SZ effect. For each cluster, the left panel shows the results for the isolated case with no EFE, while the right panels consider a basic implementation of the EFE from the large-scale structure using MOND external field strengths of 10\% (red), 30\% (blue), 50\% (orange) and 100\% (green) $a_0$. In most cases, considering a relatively strong EFE leads to physical (non-decreasing) enclosed mass profiles; exceptions are A644 and A2319 that have clear signatures of merger activity and may be out of dynamical equilibrium. }
\label{fig:EFE}
\end{figure*}

In the isolated case, we consider Eq.\,\ref{eq:mond} with
\begin{equation}
g_{\rm N}(r) = g_{\rm N, bar}(r) + g_{\rm N, mm}(r).
\end{equation}
Combining Eq.\,\ref{eq:mond} with Eq. \ref{eq:shelltheorem}, we can then infer the radial profile of the missing mass component:
\begin{equation}\label{eq:Mmm}
    M_{\rm mm}(<r) = \dfrac{r^2}{G_{\rm N}}\left[ g_{\rm obs} \mu\Big(\frac{g_{\rm obs}}{a_0}\Big) - g_{\rm N, bar}\right].
\end{equation}
The resulting mass profiles are shown in Fig.\,\ref{fig:EFE} using the so-called `simple' $\mu$ function \citep{Famaey2005}. Different interpolations function give similar results. The mass profiles are unphysical because they start to decrease after $\sim$1 Mpc and even become negative in some cases. Similar results were obtained by \citet{gravitationalxcop}. Taking the data at face value, these results would imply that MOND is not a viable theory. There are, however, three important caveats: (1) data at large radii comes from the SZ effect and there could be unknown systematics, (2) Eq.\,\ref{eq:mond} neglects the cosmic EFE from the large-scale matter distribution, and (3) hydrostatic bias may be important beyond 1 Mpc.

Regarding the first caveat, Fig.\,\ref{fig:EFE} shows that X-ray data and SZ data give consistent results in the radial range where both data are available, so we suspect that possible systematics between the two different methods play a minor role. 

Regarding the second and third caveat, both hydrostatic bias and EFE may flatten the missing mass profiles by decreasing the value of $g_{\rm obs}$, so their effects are degenerate. The EFE can be approximately taken into account using Eq.\,59 of \citet{Famaey2012}. Rearranging that equation, we derive a new equation for $M_{\rm mm}$:
\begin{equation}\label{eq:EFE}
\begin{split}
    &M_{\rm mm}(<r) = \dfrac{r^2}{G_{\rm N}} \times \\ 
    & \left\{ g_{\rm obs} \mu \Bigg(\frac{g_{\rm obs}+g_{\rm M, e}}{a_0}\Bigg) + g_{\rm M, e} \Bigg[\mu \Bigg(\frac{g_{\rm obs}+g_{\rm M, e}}{a_0}\Bigg) - \mu\Bigg(\frac{g_{\rm M, e}}{a_0}\Bigg) \Bigg] - g_{\rm N, bar} \right\},
\end{split}
\end{equation}
where $g_{\rm M, e}$ is the mean MOND gravitational field due to large-scale mass distribution in the Universe. Eq.\,\ref{eq:EFE} is mathematically equivalent to the EFE implementation in Sect.\,\ref{sec:methods} \citep[see][]{Chae2020, Chae2021}. However, we stress that $g_{\rm N, e}$ in Eq.\,\ref{eq:nuEFE} is a very different concept than $g_{\rm M, e}$ in Eq.\,\ref{eq:EFE}. The former one is the Newtonian external field from the large-scale distribution of baryons, so it can be approximately measured by summing the individual Newtonian contributions of galaxies and galaxy clusters in the nearby Universe \citep{Chae2021}. The latter one is the MOND external gravitational field, which is a difficult quantity to measure because it requires non-linear cosmological calculations of the large-scale structure in MOND. In addition, we stress that these 1D EFE formulas are approximated because they neglect the vectorial nature of the gravitational fields.

We repeated the calculation of $M_{\rm mm}$ using Eq.\,\ref{eq:EFE}, assuming the simple interpolation function and fixing $g_{\rm M, e}$ to be 10\%, 30\%, 50\% and 100\% the value of $a_0$. The results are shown in Fig.\,\ref{fig:EFE}. In three out of five cases (A1795, A2029, A2142), the EFE can flatten out the missing mass profiles, making them physical. The required external field strength is between 0.3-0.5 $a_0$, which is relatively high but still within the realm of physical reality. Higher values of $g_{\rm M, e}$ make the mass profiles increasing with radius, implying that the missing mass has not converged to a finite value, but there is a saturation effect in $M_{\rm mm}$ for $g_{\rm M, e} \gtrsim a_0$. Clearly, these basic calculations do not consider hydrostatic bias, so the values of $g_{\rm M, e}$ are as large as needed.

For A644 and A2319, there is no value of $g_{\rm M, e}$ that can avoid the decreasing profiles for the missing mass component. These two clusters are known to show signatures of merger activity (see Sect.\,\ref{sec:sample}), so the hot gas at large radii is likely out of dynamical equilibrium. If so, unphysical mass profiles would be a natural outcome of a falsifiable dynamical theory. Actually, they would testify the ability of MOND in identifying merging clusters.

\section{Baryonic scaling}\label{sec:baryonicscaling}

\begin{figure*}
\centering		 
	\includegraphics[width=4.5cm, height=4.5cm]{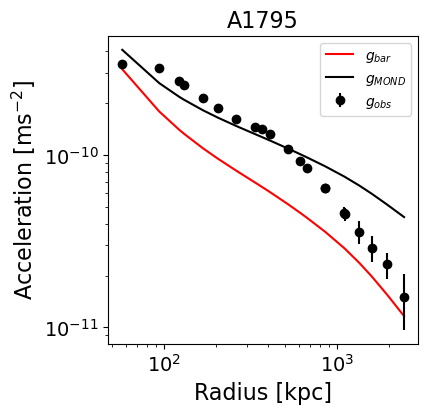}
	\includegraphics[width=4.5cm, height=4.5cm]{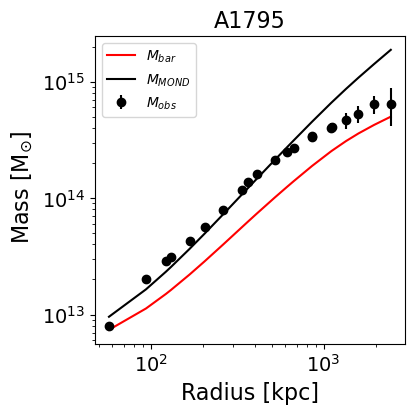}
	\includegraphics[width=4.5cm, height=4.5cm]{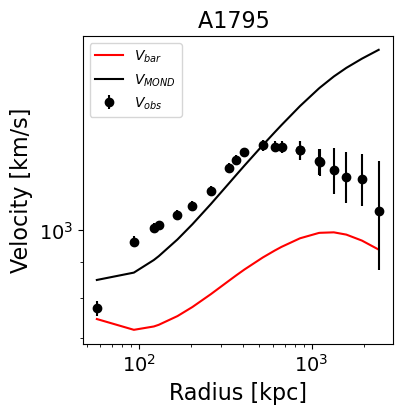}
	\includegraphics[width=4.5cm, height=4.5cm]{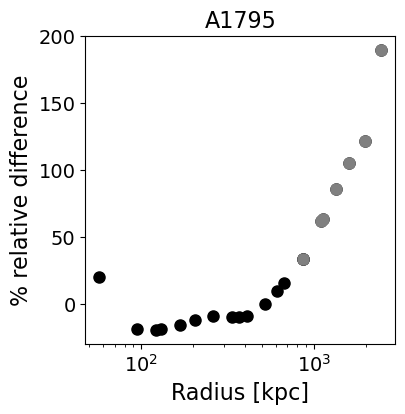}
          
	\includegraphics[width=4.5cm, height=4.5cm]{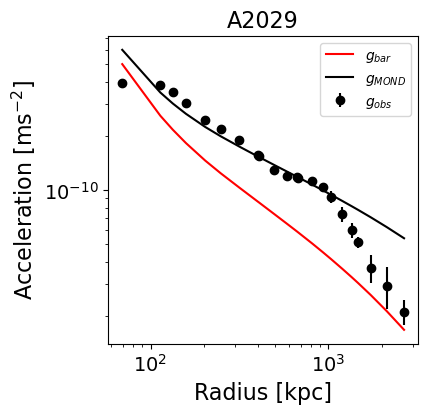}
	\includegraphics[width=4.5cm, height=4.5cm]{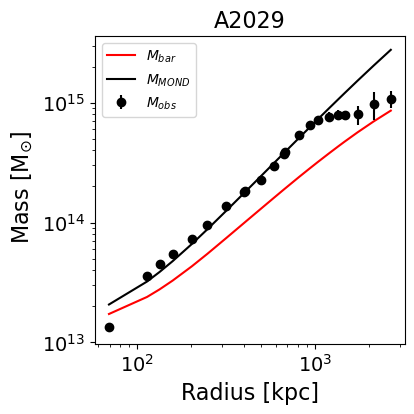}
	\includegraphics[width=4.5cm, height=4.5cm]{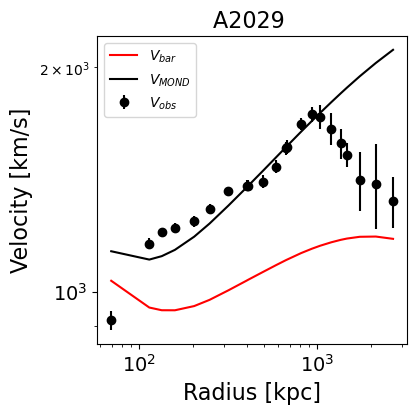}
	\includegraphics[width=4.5cm, height=4.5cm]{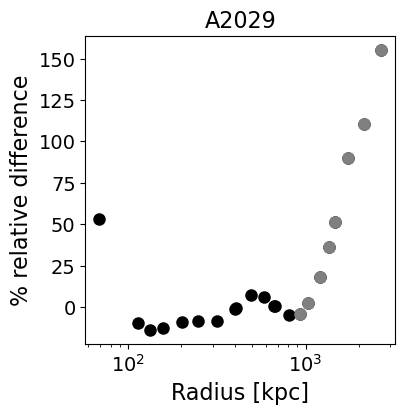}
          
	\includegraphics[width=4.5cm, height=4.5cm]{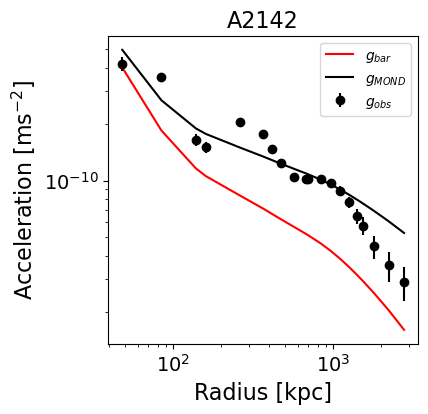}
	\includegraphics[width=4.5cm, height=4.5cm]{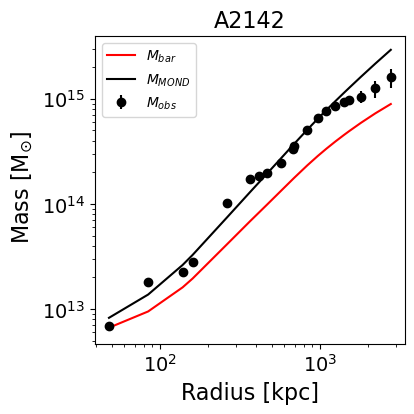}
	\includegraphics[width=4.5cm, height=4.5cm]{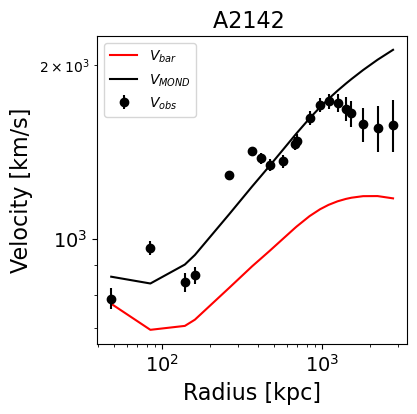}
	\includegraphics[width=4.5cm, height=4.5cm]{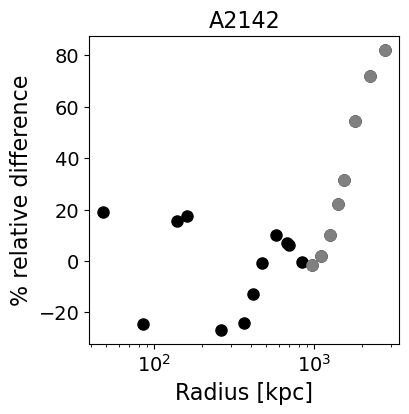}

    \includegraphics[width=4.5cm, height=4.5cm]{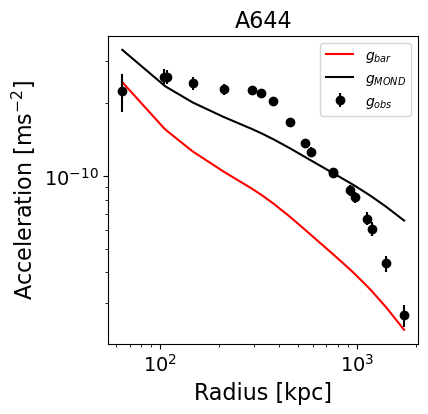}
	\includegraphics[width=4.5cm, height=4.5cm]{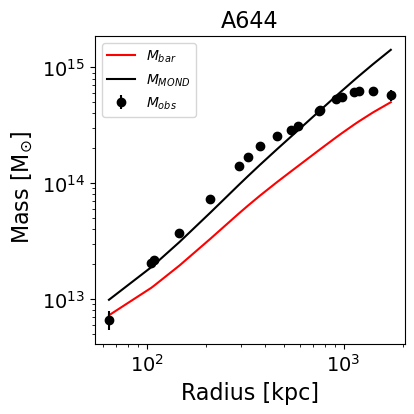}
	\includegraphics[width=4.5cm, height=4.5cm]{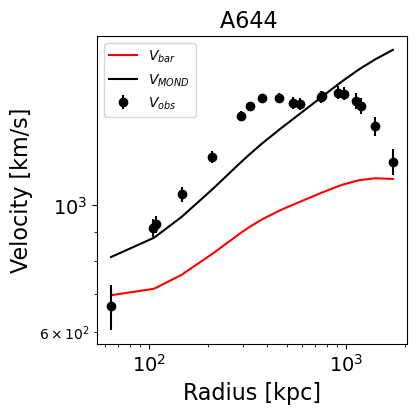}
    \includegraphics[width=4.5cm, height=4.5cm]{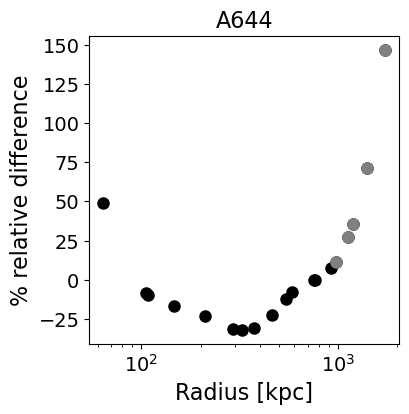}
    
	\includegraphics[width=4.5cm, height=4.5cm]{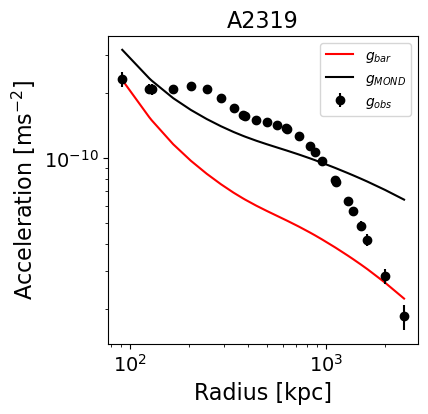}
	\includegraphics[width=4.5cm, height=4.5cm]{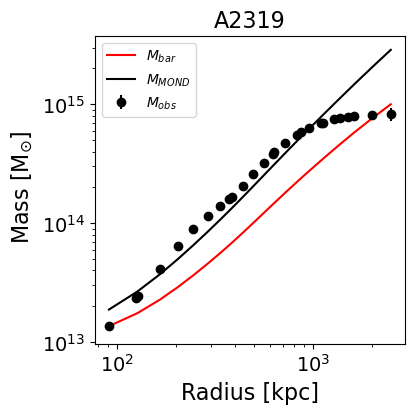}
	\includegraphics[width=4.5cm, height=4.5cm]{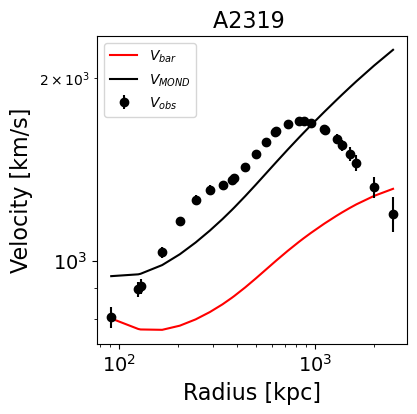}
	\includegraphics[width=4.5cm, height=4.5cm]{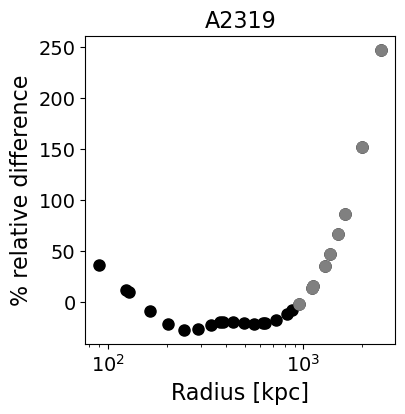}
          
	\caption{MCMC fit results for the baryonic-scaling case in isolation. Panels are the same as in Fig.\,\ref{fig:fit_efe}.}
\label{fig:fit_bary}
\end{figure*}

For completeness, we present MOND fits where the observed baryonic contribution is scaled using a single free parameter $\Upsilon_{\rm bar}$, as it is traditionally done in rotation-curve analyses of disc galaxies. In the context of galaxy clusters, these models assume that the missing matter has the same distribution as the observed matter. We consider the isolated MOND case (Eq.\,\ref{eq:nu}) and the simple interpolation function. The fit results are shown in Figure \ref{fig:fit_bary} and the best-fit values of $\Upsilon_{\rm bar}$ are provided in Table \ref{baryonicscaling}.

\begin{table}[h]
\caption{Best-fit values of $\Upsilon_{\rm bar}$ for the MOND baryonic scaling}
\label{baryonicscaling}
\begin{tabular}{ccc}
 Cluster& $ \log_{10}(\Upsilon_{\rm bar})$ & $\Upsilon_{\rm bar}$\\
\hline
A644 &
$0.684 ^{+0.00807}_{-0.00817}$ &
$4.826^ {+1.0188}_{-0.981}$
\\
 A1795&
$0.628 ^{+0.00577}_{-0.00582}$ &
$4.244 ^{+1.0134}_{-0.987}$
\\
 A2029 &
$0.551 ^{+0.00532}_{-0.00531}$&
$3.558 ^{+1.0123}_{-0.988}$
\\
 A2142 &
$0.505^{+0.00852}_ {- 0.00867}$&
$3.201^{+1.0198}_{-0.980}$
\\
A2319 &
$0.529^ {+ 0.00431}_ {-0.00437}$&
$3.800^{+1.00997}_{-0.990}$
\\
\end{tabular}
\end{table}

The best-fit values of $\Upsilon_{\rm bar}$ suggests that there must be about 3-5 times more matter than observed to reproduce the observations in MOND, but this applies only if the missing matter has the same distribution as the observed matter. The fact that the acceleration profiles are not well fit, instead, indicates that this cannot be the case: the missing mass must have a different distribution that the visible baryons, as we explicitly modeled in Sect.\,\ref{sec:fit}. In that case, the total amount of missing mass is about 0.4-1 time the amount of visible mass.

\section{Corner plots} \label{app:corner}

\begin{figure*}
\center
	   \begin{subfigure}{0.45\linewidth}
		\includegraphics[width=7cm, height=7cm]{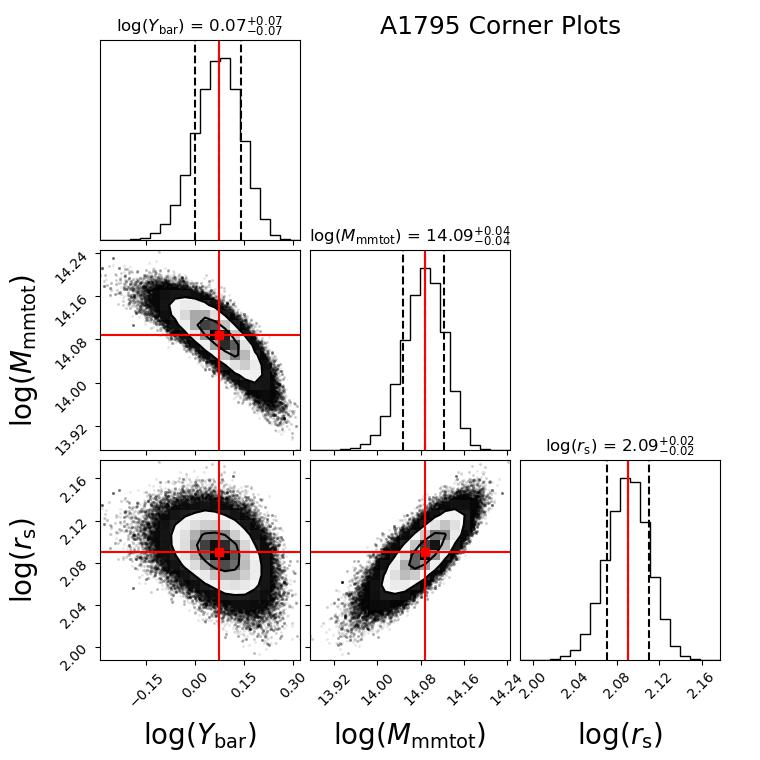}
        \vfill
	    \end{subfigure}\\
     
        \begin{subfigure}{0.45\linewidth}
		\includegraphics[width=7cm, height=7cm]{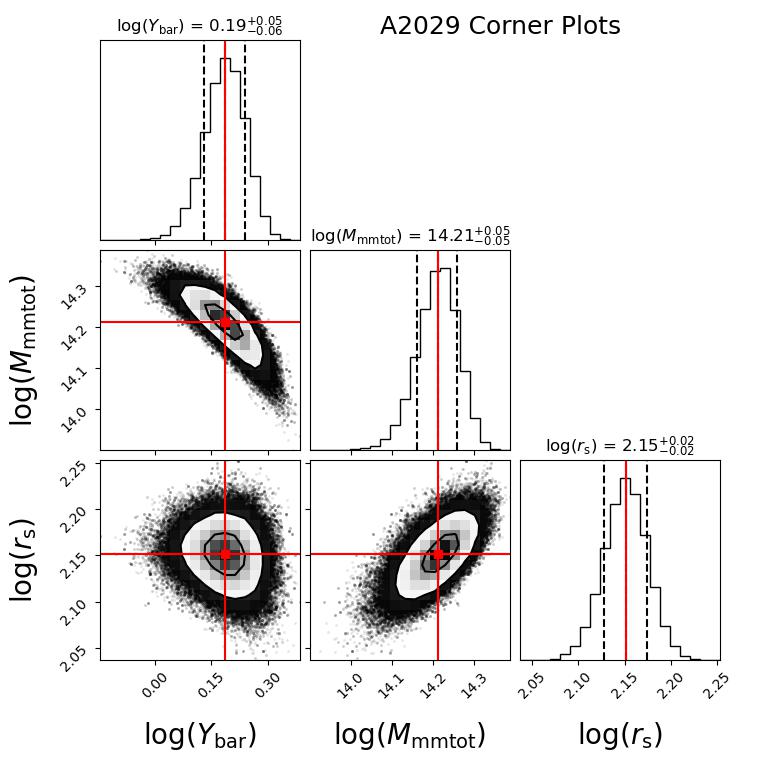}
	   \end{subfigure}
	   \begin{subfigure}{0.45\linewidth}
		\includegraphics[width=7cm, height=7cm]{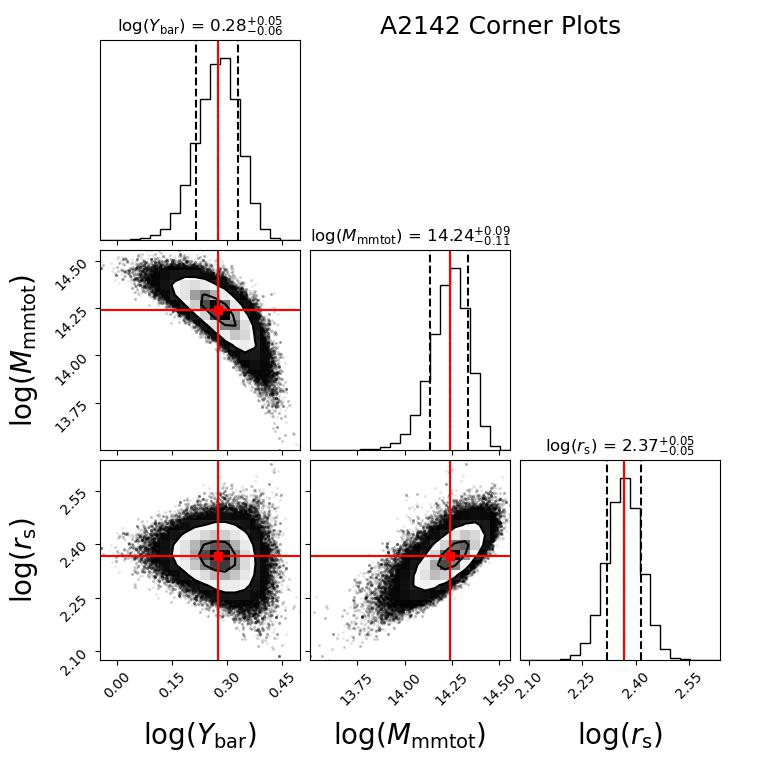}
        \vfill
	    \end{subfigure}

    \begin{subfigure}{0.45\linewidth}
		\includegraphics[width=7cm, height=7cm]{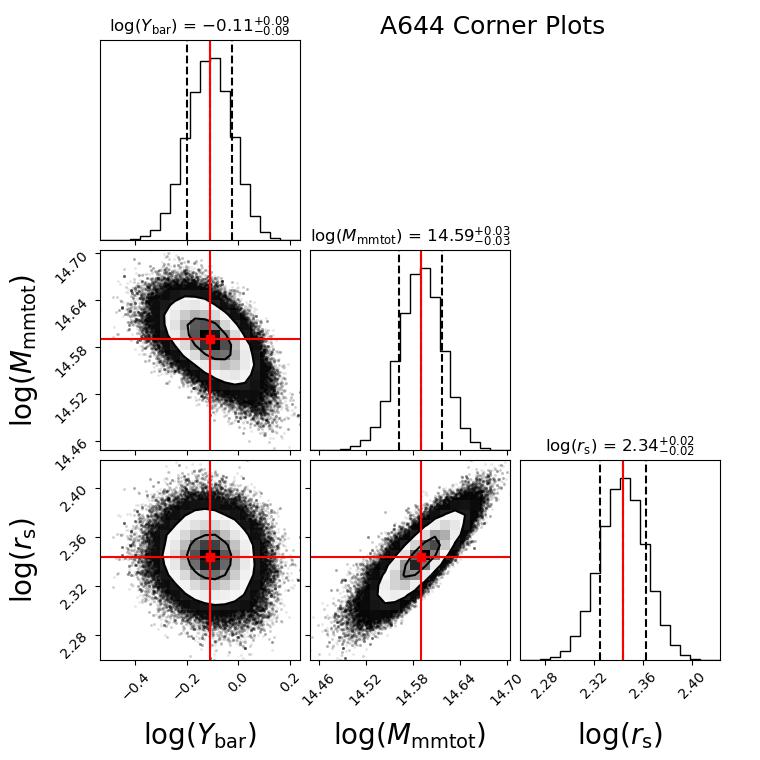}
	   \end{subfigure}
	\begin{subfigure}{0.45\linewidth}
		\includegraphics[width=7cm, height=7cm]{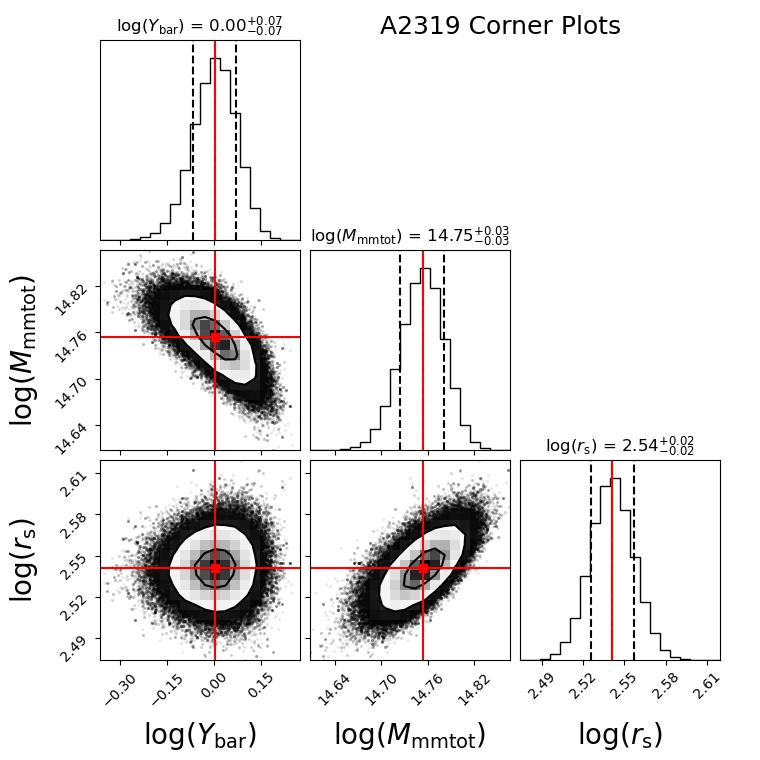}
	    \end{subfigure}
       
	    \caption{Posterior probability distributions from MCMC fits in the isolated case. See Appendix \ref{app:corner} for details.} \label{fig:corner_noEFE}
\end{figure*}

\begin{figure*}
\center
	   \begin{subfigure}{0.45\linewidth}
		\includegraphics[width=7cm, height=7cm]{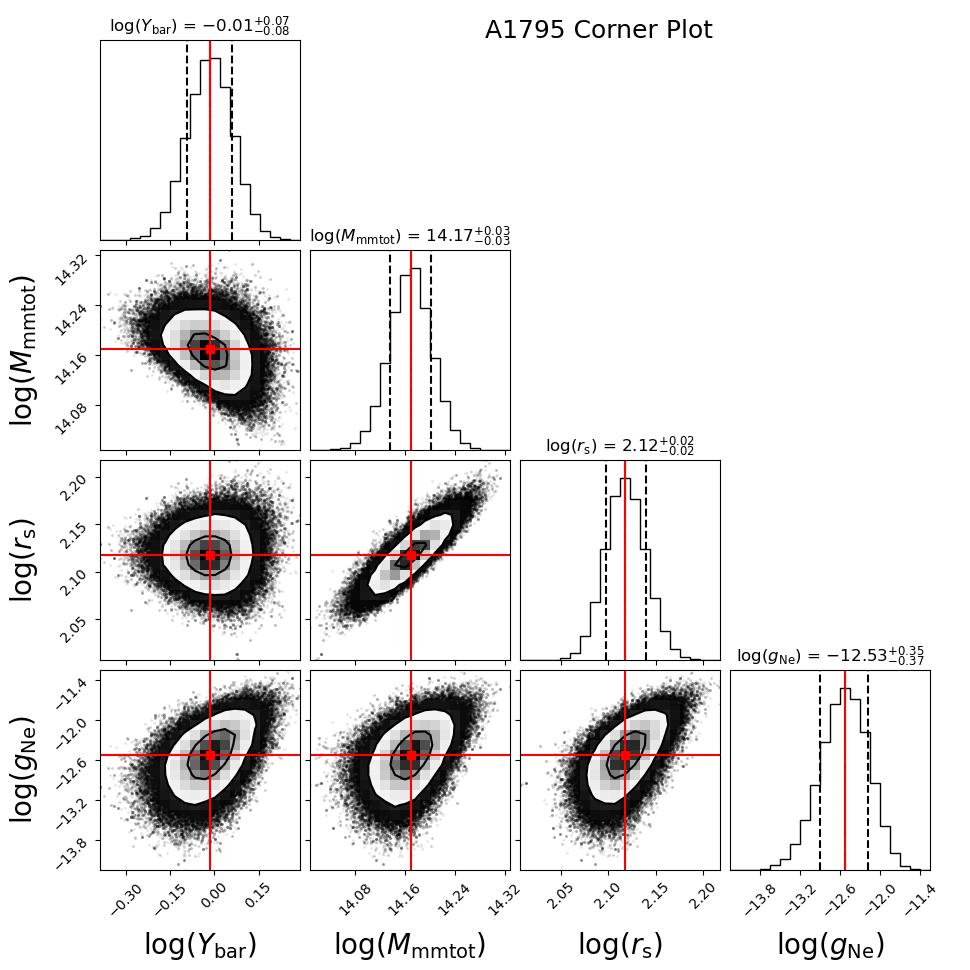}
        \vfill
	    \end{subfigure}\\

        \begin{subfigure}{0.45\linewidth}
		\includegraphics[width=7cm, height=7cm]{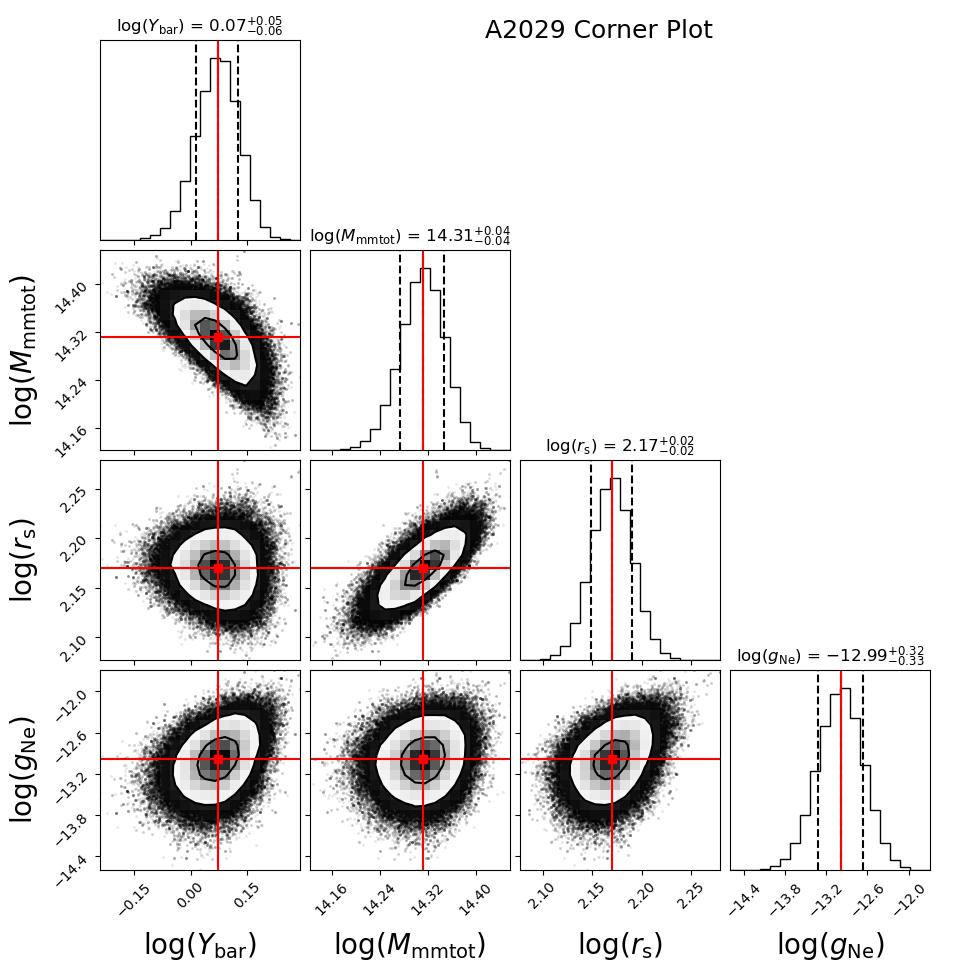}
	   \end{subfigure}
	   \begin{subfigure}{0.45\linewidth}
		\includegraphics[width=7cm, height=7cm]{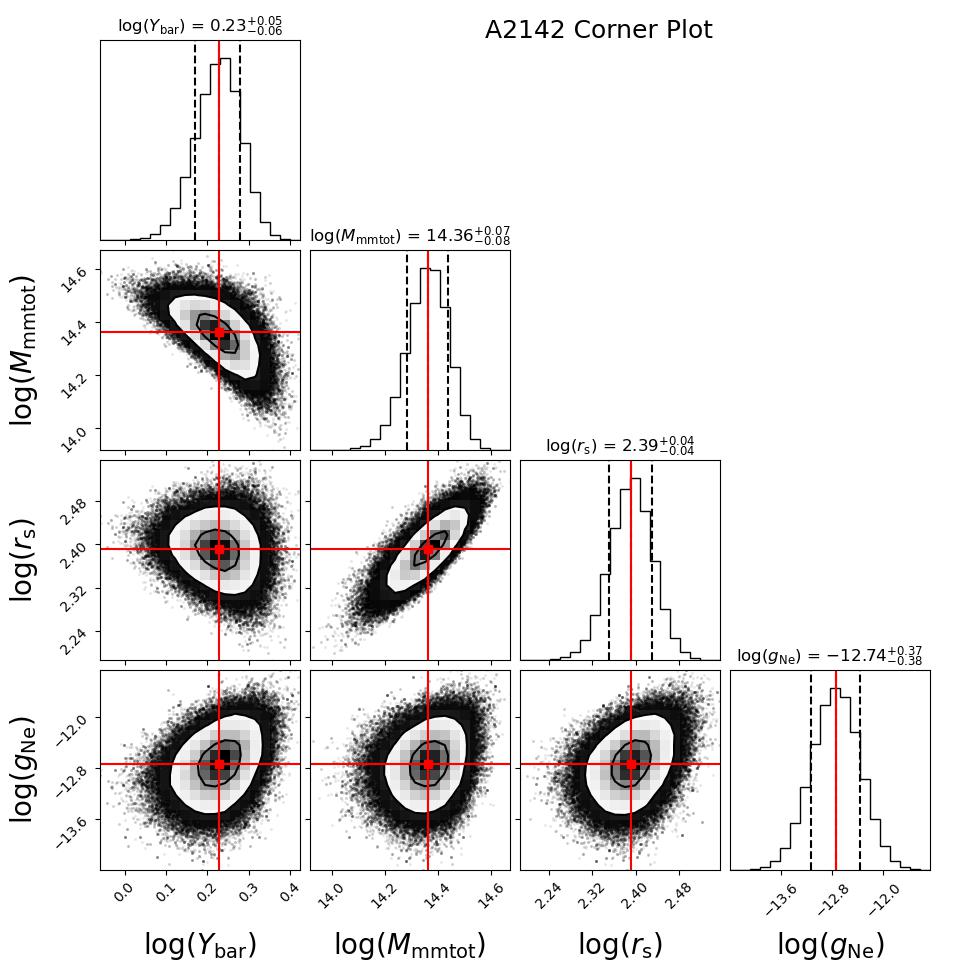}
        \vfill
	    \end{subfigure}

    \begin{subfigure}{0.45\linewidth}
		\includegraphics[width=7cm, height=7cm]{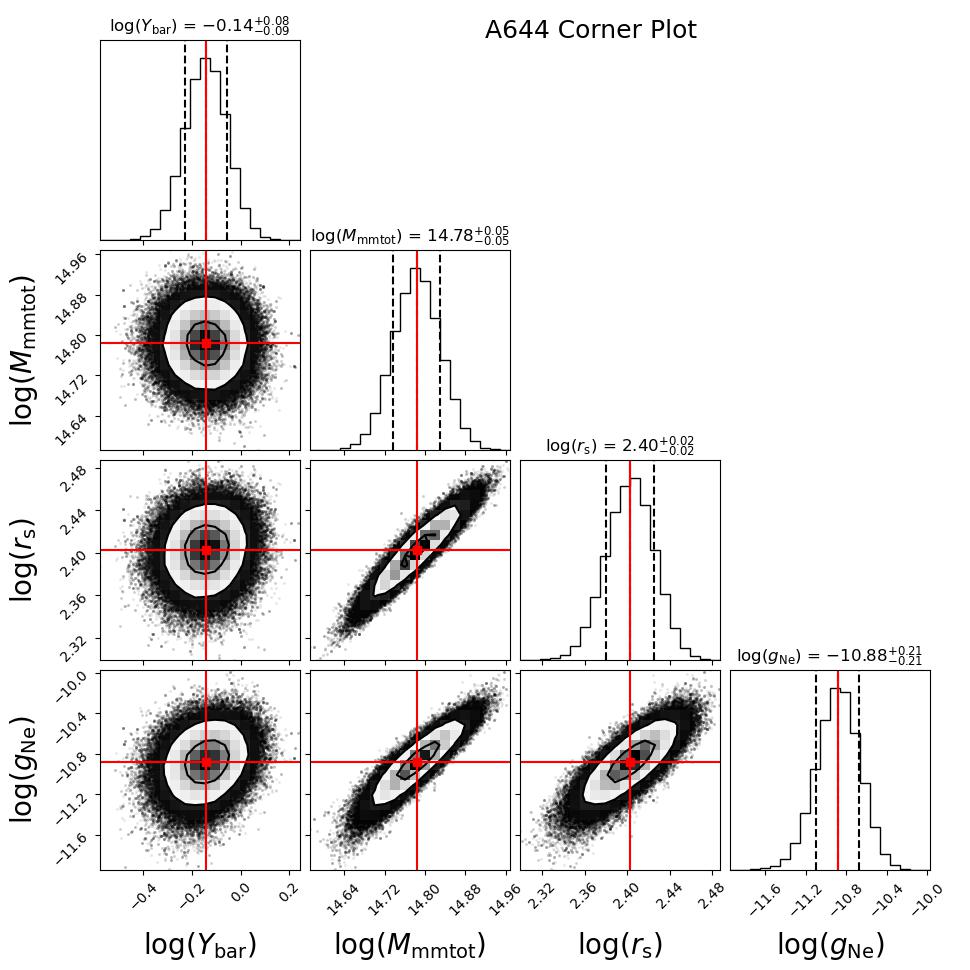}
	   \end{subfigure}
	\begin{subfigure}{0.45\linewidth}
		\includegraphics[width=7cm, height=7cm]{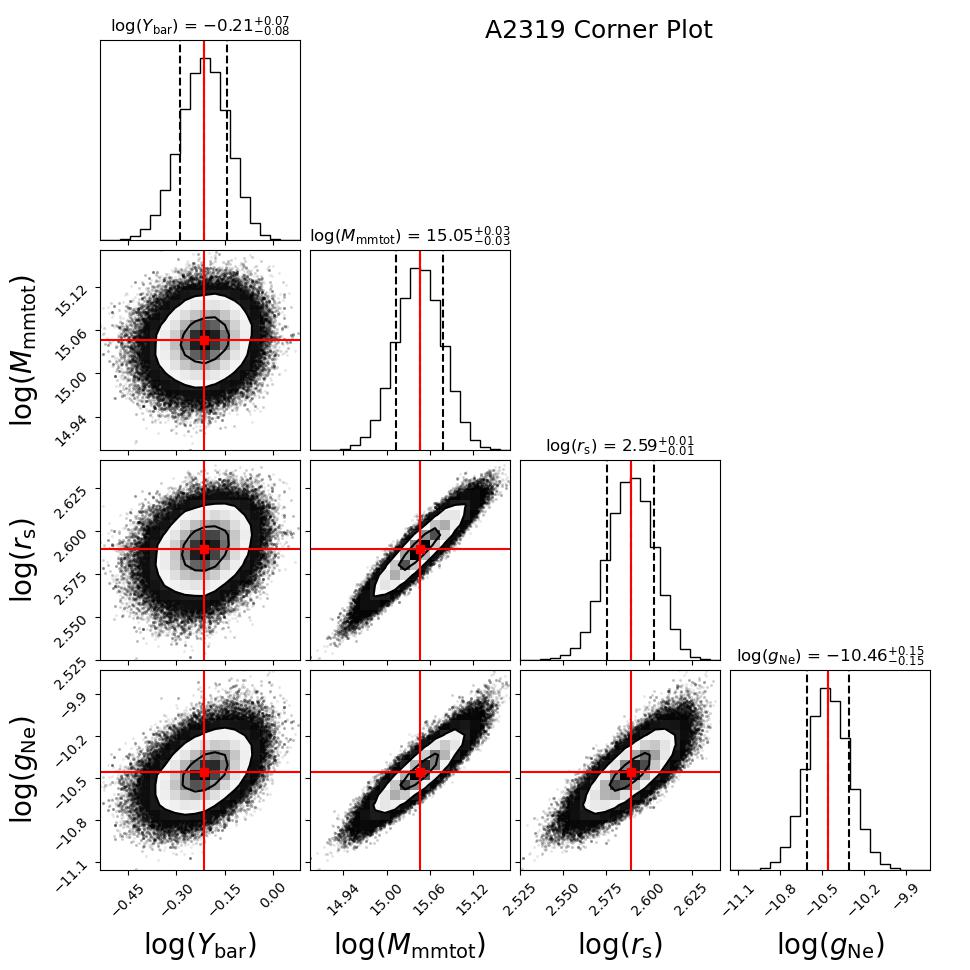}
	    \end{subfigure}
       
	    \caption{Posterior probability distributions from MCMC fits in the EFE case. See Appendix \ref{app:corner} for details.}   \label{fig:corner_EFE}
\end{figure*}

Figures\,\ref{fig:corner_noEFE} and \ref{fig:corner_EFE} show `corner plots' from MCMC fits in the isolated and EFE cases, respectively. For each cluster, the various panels show the posterior probability distribution of pairs of fitting parameters, and the marginalised probability distribution of each fitting parameter. In the inner panels, individual MCMC samples outside the 2$\sigma$ confidence region are shown with black dots, while binned MCMC samples inside the 2$\sigma$ confidence region are shown by a greyscale; black contours correspond to the 1$\sigma$ and 2$\sigma$ confidence regions; the red squares and red solid lines show median values. In the outer panels (histograms), solid and dashed lines correspond, respectively, to the median and $\pm1\sigma$ values, which are reported at the top of the panel. In general, the posterior probability distributions are well-behaved and show clear peaks, indicating that the fitting quantities and their uncertainties are well measured.

\end{appendix}
\end{document}